\documentclass[12pt]{article}
\usepackage{amsmath,amssymb,amsthm,booktabs}
\usepackage{mathrsfs}
\usepackage{geometry}
\usepackage{pifont}
\usepackage[OT1]{fontenc}
\usepackage[utf8]{inputenc}
\usepackage[colorlinks,citecolor=blue,urlcolor=blue]{hyperref}
\usepackage{txfonts}
\usepackage{indentfirst}
\usepackage{multirow}
\usepackage{float,subfig}

\usepackage{bm}
\usepackage{euscript}
\usepackage{graphicx}
\usepackage{multicol}
\usepackage[usenames,dvipsnames,svgnames,table]{xcolor}

\usepackage[round]{natbib}
\usepackage[ruled]{algorithm2e}
\usepackage{pgfplots}
\pgfplotsset{compat=1.17}

\numberwithin{equation}{section} \theoremstyle{plain}
\newtheorem{theorem}{Theorem}[section]

\newtheorem{corollary}{Corollary}[section]

\newtheorem{example}{Example}[section]
\newtheorem{definition}{Definition}[section]
\newtheorem{remark}{Remark}[section]

\renewcommand \arraystretch{1.3}
\begin{document}

\newcommand{\gai}[1]{{#1}}


\makeatletter
\def\ps@pprintTitle{%
  \let\@oddhead\@empty
  \let\@evenhead\@empty
  \let\@oddfoot\@empty
  \let\@evenfoot\@oddfoot
}
\makeatother

\newcommand\tabfig[1]{\vskip5mm \centerline{\textsc{Insert #1 around here}}  \vskip5mm}

\vskip2cm

\title{Robust asset pricing and superhedging duality under model uncertainty with and without short-sale constraints}
\author{Wenqing Zhang \thanks{Corresponding author, College of Mathematical Sciences, China Three Gorges University, PR China, (wqzhang1025@163.com).}
\quad Shuzhen Yang \thanks{Zhong Tai Securities Institute for Financial Studies, Shandong University, PR China, (yangsz@sdu.edu.cn). This work was supported by the National Key R\&D program of China (Grant No.2018YFA0703900,\ ZR2019ZD41), National Natural Science Foundation of China (Grant No.11701330), and Taishan Scholar Talent Project Youth Project.}
}
\date{}
\maketitle

\begin{abstract}
We study asset pricing and hedging under model uncertainty in discrete time and finite states. For the single-period model, we characterize no-arbitrage by the existence of strictly positive martingale measures without short-sale restrictions and strictly positive supermartingale measures under short-sale prohibitions. In the constrained setting, we further consider a family of probability measures and introduce weak and strong risk-neutral nonlinear expectations. Their existence provides a necessary condition and a sufficient condition for no-arbitrage, respectively. We extend these results to multi-period markets, derive risk-neutral prices for replicable claims, and establish superhedging duality over the complete martingale and supermartingale families. A numerical illustration calibrated to monthly S\&P 500 returns implements the pricing and superhedging framework for synthetic put and call options. The optimal values of the primal and complete dual problems agree to numerical precision, as predicted by the superhedging duality theorems. The results further show that short-sale prohibitions substantially increase put superhedging costs, while call costs are unchanged.
\end{abstract}

\noindent KEYWORDS: Discrete time and states; Sublinear expectation; Short-sales prohibitions; Asset pricing; Superhedging duality

\section{Introduction}
\label{sec:introduce}

The fundamental theorem of asset pricing is a cornerstone of mathematical finance, linking the absence of arbitrage to the existence of an appropriate probability measure under which discounted asset prices satisfy a martingale condition. It has been developed extensively across a broad range of market settings \citep{Delbaen1994,Kuhn2025,Lyasoff2014}. Discrete-time models are particularly natural when portfolio positions are revised at specified trading dates and are also well suited to numerical implementation \citep{Bertsimas00}. Accordingly, a substantial literature has studied asset pricing and hedging in discrete time \citep{Brennan79,Perrakis84,Bielecki15,Burzoni16}.

Portfolio constraints further modify this classical characterization. When short selling is prohibited, admissible holdings in risky assets are constrained to be nonnegative, and supermartingale measures arise naturally in the corresponding dual formulation \citep{Pulido14,Coculescu19}. From an economic perspective, short-sale constraints may prevent pessimistic views from being fully incorporated into market prices and may thereby contribute to overvaluation \citep{Battalio11}. They also make the pricing and hedging of derivatives more involved \citep{He20}.

Model uncertainty constitutes another departure from the classical single-prior setting. It is commonly represented by a family $\mathcal{P}$ of plausible probability measures rather than by a single reference measure. The theory of nonlinear expectations developed by Peng, and particularly its sublinear form, provides a natural framework for aggregating the resulting expectations \citep{Peng1997,Peng2004,Peng2006,Peng2008,Peng2019}. It has found broad applications to mean and volatility uncertainty in financial markets \citep{EJ13,EJ14,Peng2022,Peng2023}. In this setting, the classical fundamental theorem of asset pricing and superhedging duality require robust reformulation. For a nondominated discrete-time market governed by $\mathcal{P}$, \citet{Bouchard15} established a robust fundamental theorem of asset pricing and a superhedging duality under quasi-sure no-arbitrage. Under this condition, the associated family $\mathcal{Q}$ of martingale measures and $\mathcal{P}$ have the same polar sets, and the minimum superhedging price is the supremum of expectations over $\mathcal{Q}$. Related robust and pointwise formulations have been developed in \citet{Burzoni19,Obloj21}.

Despite these advances, the joint treatment of model uncertainty and short-sale prohibitions remains comparatively limited, particularly in a framework that covers both single- and multi-period pricing and superhedging and is amenable to direct numerical implementation. We therefore work on a finite state space in the single-period model and on a finite event tree in the multi-period model. This specification makes the separation arguments and the primal-dual optimization problems explicit and corresponds directly to the trinomial construction used in our real-data numerical illustration. Within this framework, we investigate asset pricing and superhedging under model uncertainty, both with and without short-sale prohibitions.

For the single-period securities model, we consider a finite state space and use a family of probability measures $\mathcal P$ to represent model uncertainty. The collective-support condition $\sup_{P\in\mathcal P}P(\omega)>0$ for every state $\omega$ eliminates nonempty $\mathcal P$-polar sets; consequently, quasi-sure inequalities are equivalent to statewise inequalities in the present framework. 
In the unrestricted market, no-arbitrage is equivalent to the existence of a strictly positive martingale measure. Under short-sale prohibitions, the corresponding condition is the existence of a strictly positive supermartingale measure. 
Since an incomplete market generally admits multiple risk-neutral probability measures, we introduce a family of probability measures $\mathcal Q$ and formulate weak and strong risk-neutral nonlinear expectation conditions.
We show that the weak condition is necessary for no-arbitrage, whereas the strong condition is sufficient. The upper expectation over a family satisfying the strong condition provides a conservative valuation.

We next extend these results to a multi-period finite event tree. Strictly positive local kernels are pasted along the tree to construct global pricing measures, and rectangularity is imposed when a family of conditional models is evaluated recursively. The unrestricted market is characterized by martingale kernels, whereas the short-sale-constrained market is characterized by supermartingale kernels; the weak and strong nonlinear-expectation results extend accordingly. 
For replicable claims, the usual discounted risk-neutral valuation is recovered. For general contingent claims, we establish exact superhedging dualities: the minimum superhedging price is the supremum of discounted expected payoffs over the complete martingale family in the unrestricted market and over the complete supermartingale family under short-sale prohibitions. 
These dualities are implemented by nodewise linear programming and backward induction, which jointly produce the minimum initial capital, an optimal trading strategy, and the associated dual kernels. A valuation based on a prescribed ambiguity family is therefore distinguished from the exact superhedging price based on the complete dual family.

To examine the quantitative implications of this distinction, we provide a rolling real-data-calibrated numerical application using monthly S\&P 500 price-index observations from January 1990 through August 2026. 
At each of 317 valuation dates from February 2000 to June 2026, a two-period trinomial tree is constructed using only information available before the valuation date. The probability bounds are calibrated from martingale kernels estimated over 36-, 60-, and 120-month windows, and synthetic two-month at-the-money European put and call options are valued under a single-model benchmark, data-supported martingale and supermartingale families, and the corresponding complete dual families. 
The separately solved primal and complete dual problems agree to numerical precision. Expanding the probability bounds widens the data-supported valuation intervals, while their upper endpoints generally remain below the exact superhedging prices.
The short-sale prohibition has a large, payoff-dependent effect: the mean put superhedging cost rises from $3.583\%$ to $13.648\%$ of the contemporaneous index level because the unrestricted put hedge uses negative stock positions, whereas the mean call cost remains $3.913\%$ in both trading regimes because the optimal call positions are nonnegative. 
In the subsequent-path evaluation, hedging deficits occur only in windows containing returns outside the calibrated range. The numerical results thus separate the effect of probability ambiguity on valuation from the effect of trading constraints on attainable hedges and superhedging costs.

The main contributions of this paper are fourfold.

(i). We develop a finite-state asset-pricing framework under model uncertainty for both single- and multi-period markets. The full support condition makes the robust no-arbitrage condition statewise and yields a transparent characterization in terms of strictly positive martingale measures, with the multi-period result constructed from local conditional kernels.

(ii). We incorporate short-sale prohibitions into the framework and show that strictly positive supermartingale measures replace martingale measures in the no-arbitrage characterization. For a family of probability measures $\mathcal Q$, we formulate weak and strong risk-neutral nonlinear expectations and establish that: the existence of a weak risk-neutral nonlinear expectation is necessary for no-arbitrage, while a strong risk-neutral nonlinear expectation is sufficient. 

(iii). We derive pricing, hedging, and superhedging results for contingent
claims. Replicable claims admit risk-neutral valuation, while the exact
superhedging prices are characterized by the complete martingale and
supermartingale dual families in the unrestricted and short-sale-constrained
markets, respectively. A unified backward procedure based on nodewise linear
programs makes the duality operational and recovers both optimal values and
trading strategies.

(iv). We implement the theoretical framework in a rolling numerical application calibrated to real S\&P 500 data. 
The analysis provides a numerical consistency check on the superhedging dualities by independently solving the primal and complete dual problems, examines how model uncertainty and short-sale prohibitions affect put and call valuations and superhedging costs, and evaluates the resulting hedging strategies along subsequent observed paths.

The remainder of this paper is organized as follows. 
Section \ref{sec:single} studies the single-period market and establishes the no-arbitrage characterizations without and with short-sale prohibitions.
Section \ref{sec:multi} extends the analysis to a multi-period finite event tree. 
Section \ref{sec:hedging} derives the replication and superhedging results and presents the backward linear-programming algorithm. 
Section \ref{sec:empirical} reports the rolling real-data-calibrated numerical application and the subsequent-path hedge evaluation. 
Section \ref{sec:conclude} concludes.

\section{Single-period asset pricing}
\label{sec:single}

For a single-period securities model, we consider the finite state space $\Omega=\{\omega_k\}_{k=1}^{N}$.
The market consists of $M$ risky securities $S_m$, $m=1,2,\cdots,M$, and a risk-free bond $S_0$.
Let $\{S_i(t),\ t=0,1,\ i=1,2,\cdots,M\}$ denote the price processes.
The initial prices of the risky securities are positive and known at $t=0$, while their values at $t=1$ are random variables.
To describe model uncertainty, we introduce a family of probability measures $\mathcal{P}$ on $\Omega$, see \citet{YZ26}.
We assume $\sup_{P\in\mathcal{P}}P(\omega)>0$ for all $\omega\in\Omega$.
Without loss of generality, we take $S_0(0)=1,\ S_0(1)=1+r$, where $r$ is the deterministic single-period interest rate satisfying $1+r>0$.
The quantity $S_0(1)=1+r$ is the gross risk-free return, and $1/S_0(1)$ is the discount factor over the period.
The discounted price processes are defined by
\begin{equation}
	\label{eq:discount price}
	S_i^{*}(t)=S_i(t)/S_0(t),\quad t=0,1,\ i=1,2,\cdots,M.
\end{equation}

At time $0$, an investor chooses a portfolio with trading strategy $h_m$, $m=0,1,\cdots,M$, where $h_m$ denotes the number of units of the $m$-th asset held from $t=0$ to $t=1$.
Let $\{V_t,t=0,1\}$ denote the total value of the portfolio,
$$
V_t=h_0S_0(t)+\sum_{m=1}^M h_mS_m(t),\quad t=0,1.
$$
The gain on the $m$-th risky security is $h_m\Delta S_m=h_m[S_m(1)-S_m(0)]$, and the total gain $G$ of the portfolio is
$$
G=h_0r+\sum_{m=1}^{M} h_m\Delta S_m.
$$
We assume that there is no withdrawal or addition of funds within the investment horizon, and then $V_1=V_0+G$.
Based on the discounted asset price $S^{*}(t)$ in equation \eqref{eq:discount price}, we define the discounted value process by $V_{t}^{*}=V_t/S_0(t)$ and the discounted gain becomes $G^{*}=V_1^{*}-V_0^{*}$, that is,
\begin{align}
	\label{eq:value discount}
	V_t^{*}&=h_0+\sum_{m=1}^{M}h_m S_{m}^{*}(t),\quad t=0,1,\\
	\label{eq:gain discount}
	G^{*}&=V_1^*-V_0^*=\sum_{m=1}^{M} h_m \Delta S_{m}^{*},
\end{align}
where $\Delta S_{m}^{*}=S_m^*(1)-S_m^*(0)$.

\subsection{Asset pricing under uncertainty}

The fundamental theorem of asset pricing plays a critical role in the securities model by characterizing the relationship between no-arbitrage and risk neutral probability measures.
In a classical finite-state market, an arbitrage opportunity has the following properties: (i) $V_0^*=0$; (ii) $V_1^*(\omega)\geq 0$ for all $\omega\in\Omega$ and $E_P[V_1^*]>0$, where $P$ is a probability measure satisfying $P(\omega)>0$ for every $\omega\in\Omega$, see \citet{Kwok08}.
Under model uncertainty, a single probability measure is replaced by a family of probability measures $\mathcal P$, leading to a family of linear expectations $\{E_P[\cdot]:P\in\mathcal P\}$.
It is then natural to consider the lower and upper expectations
$\inf_{P\in\mathcal P}E_P[V_1^*]$ and
$\sup_{P\in\mathcal P}E_P[V_1^*]$.
For a nonnegative terminal payoff, the condition
$\inf_{P\in\mathcal P}E_P[V_1^*]>0$
requires the expected payoff to be uniformly positive over all possible models.
This condition is stronger than what is needed for an arbitrage opportunity, which only requires a strictly positive expected payoff under at least one possible model.
Therefore, we consider the upper expectation
$\sup_{P\in\mathcal P}E_P[V_1^*]$.
Together with the requirement that the terminal payoff is nonnegative in every state, this leads to the following definition of arbitrage under model uncertainty.

\begin{definition}[\textbf{Arbitrage under model uncertainty}]
	\label{de:arbitrage uncertainty}
	A trading strategy $h$ is said to be an arbitrage opportunity if the discounted value of the portfolio satisfies:
	
	(i)$\ V_{0}^{*}=0$;
	
	(ii)$\ V_{1}^{*}(\omega)\ge 0, \ \forall \omega\in\Omega\ $ and $\ \sup_{P\in\mathcal{P}} E_{P}[V_{1}^{*}]>0$,
	
	\noindent where $E_P[\cdot]$ denotes the expectation under a possible probability measure $P\in\mathcal{P}$.
	The market satisfies no arbitrage under model uncertainty, denoted by $NA(\mathcal P)$, if no such strategy exists.
\end{definition}

\begin{remark}
	\label{re:arbitrage}
	Recall that a set $A\subseteq\Omega$ is called $\mathcal P$-polar if
	$
	\sup_{P\in\mathcal P}P(A)=0.
	$
	A property is said to hold $\mathcal P$-quasi-surely ($\mathcal P$-q.s.) if it holds outside a $\mathcal P$-polar set.
	Under the support condition
	$
	\sup_{P\in\mathcal P}P(\omega)>0,
	\ \forall\,\omega\in\Omega,
	$
	there are no nonempty $\mathcal P$-polar sets.
	Indeed, if $A\subseteq\Omega$ is nonempty and $\omega\in A$, then
	$
	\sup_{P\in\mathcal P}P(A)
	\geq
	\sup_{P\in\mathcal P}P(\omega)
	>0.
	$
	Hence, in the present discrete-state setting, $\mathcal P$-q.s. properties are equivalent to pointwise properties.
	
	Since $V_1^*\geq0$, the same support condition implies that
	$
	\sup_{P\in\mathcal P}E_P[V_1^*]>0
	$
	if and only if $V_1^*(\omega)>0$ for at least one state $\omega\in\Omega$.
	Therefore, Definition \ref{de:arbitrage uncertainty} is equivalent both to the $\mathcal P$-quasi-sure notion of arbitrage in Definition 1.1 of \citet{Bouchard15} and to the standard statewise notion of arbitrage in the present discrete-state market.
	
	Moreover, for a zero-cost trading strategy, $V_0^*=0$ and hence $V_1^*=G^*$ by equation \eqref{eq:gain discount}.
	Thus, Definition \ref{de:arbitrage uncertainty} can equivalently be formulated in terms of the discounted gain as
	$
	G^*(\omega)\geq0,
	\ \forall\,\omega\in\Omega,
	$
	and
	$
	\sup_{P\in\mathcal P}E_P[G^*]>0.
	$
\end{remark}

Definition \ref{de:arbitrage uncertainty} describes a trading strategy with zero initial investment and a nonnegative terminal payoff in every possible state, which is strictly positive under at least one possible model.
To characterize the absence of such arbitrage opportunities, we consider a probability measure under which the expected discounted gain of each risky security is zero.
In a finite-state market, such a measure is called a risk neutral probability measure, as in \citet{Kwok08}.
It is required to satisfy the following conditions:

(i) $\ Q(\omega)>0, \quad \text{for all}\ \omega\in\Omega$;

(ii) $\ E_{Q}[\Delta S_m^*]=0, \quad m=1,\cdots,M$,

\noindent where condition $E_{Q}[\Delta S_m^*]=0$ is equivalent to
\begin{equation}
	\label{eq:risk neutral1}
	S_m^*(0)=\sum_{k=1}^{N}Q(\omega_k)S_m^*(1;\omega_k), \quad m=1,\cdots,M.
\end{equation}

Under the support condition on $\mathcal P$, the no-arbitrage condition in Definition \ref{de:arbitrage uncertainty} is equivalent to the usual statewise no-arbitrage condition.
Consequently, the classical characterization of no-arbitrage by risk neutral probability measures remains valid in the present model.
We state this result in the following theorem, which will serve as a basis for the subsequent analysis.

\begin{theorem}[\textbf{Fundamental theorem of asset pricing under model uncertainty}]
	\label{theo:fundamental}
	No-arbitrage condition under model uncertainty holds if and only if there exists a risk neutral probability measure $Q$.
\end{theorem}

\begin{proof}
We first prove the "if" direction. Suppose that there exists a strictly positive risk-neutral probability measure $Q$. If $V_1^*(\omega)\ge0$ for all $\omega\in\Omega$ and $\sup_{P\in\mathcal P}E_P[V_1^*]>0$, then $V_1^*(\omega_k)>0$ for at least one state. Since $Q(\omega_k)>0$ for every state,
\[
V_0^*=h_0+\sum_{m=1}^{M}h_mE_Q[S_m^*(1)]=E_Q[V_1^*]>0,
\]
which contradicts $V_0^*=0$. Hence no arbitrage in the sense of Definition \ref{de:arbitrage uncertainty} exists.

Now, we prove the "only if" direction. Suppose that the no-arbitrage condition under model uncertainty holds. Define the set of attainable discounted gains by
\[
U_1
:=
\left\{
\left(
G^*(\omega_1),\ldots,G^*(\omega_N)
\right)^\top
:
G^*
=
\sum_{m=1}^{M}h_m\Delta S_m^*,
\quad
h\in\mathbb R^M
\right\},
\]
and let
\[
U_2:=\mathbb R_+^N.
\]
Since unrestricted trading strategies are closed under addition and scalar multiplication, $U_1$ is a linear subspace of $\mathbb R^N$, and is therefore closed and convex. Moreover, by the support assumption
$
\sup_{P\in\mathcal P}P(\omega)>0,\ \forall \omega\in\Omega,
$
the no-arbitrage condition is equivalent to
\[
U_1\cap U_2=\{0\}.
\]
Indeed, if $y\in U_1\cap U_2$ and $y\neq0$, then $y$ is nonnegative in every state and strictly positive in at least one state, say $\omega_{k_0}$. By the support assumption, there exists $P_0\in\mathcal P$ such that $P_0(\omega_{k_0})>0$. Hence,
$
E_{P_0}[y]>0,
$
and the corresponding zero-cost trading strategy constitutes an arbitrage, which is a contradiction.
Consider the compact convex set
\[
D
:=
\left\{
x\in\mathbb R_+^N:
\sum_{k=1}^{N}x_k=1
\right\}.
\]
Since $D\subset U_2\setminus\{0\}$, the condition $U_1\cap U_2=\{0\}$ implies that
$
U_1\cap D=\varnothing.
$
Therefore, by the finite-dimensional strong separating-hyperplane theorem, there exists a nonzero vector
$
q=(q_1,\ldots,q_N)^\top\in\mathbb R^N
$
such that
\begin{equation}
	\label{eq:theo-sup<inf}
	\sup_{y\in U_1}q^\top y < \inf_{x\in D}q^\top x.
\end{equation}
Since $U_1$ is a cone containing the origin, we must have
$
q^\top y\leq0
$ for all $y\in U_1$.
Indeed, if $q^\top y_0>0$ for some $y_0\in U_1$, then $\lambda y_0\in U_1$ for every $\lambda>0$, and consequently
\[
\sup_{y\in U_1}q^\top y
\geq
\lambda q^\top y_0
\longrightarrow+\infty,
\]
which contradicts the separating inequality \eqref{eq:theo-sup<inf}. Since $0\in U_1$, it follows that
$
\sup_{y\in U_1}q^\top y=0,
$
and hence
\[
\inf_{x\in D}q^\top x>0.
\]
For each $k=1,\ldots,N$, the standard unit vector $e_k$ belongs to $D$. Therefore,
$
q_k=q^\top e_k>0,
\ 
k=1,\ldots,N,
$
and thus
$
q\in\mathbb R_{++}^N.
$
Since $U_1$ is a linear subspace, $y\in U_1$ implies $-y\in U_1$. Applying the inequality $q^\top y\leq0$ to both $y$ and $-y$, we obtain
\[
q^\top y=0,
\qquad
\forall\,y\in U_1.
\]
Define a probability measure $Q$ on $\Omega$ by
\[
Q(\omega_k)
:=
\frac{q_k}{\sum_{j=1}^{N}q_j},
\qquad
k=1,\ldots,N.
\]
Since $q_k>0$ for every $k$, we have
$
Q(\omega_k)>0,
\ 
k=1,\ldots,N,
$
and
$
\sum_{k=1}^{N}Q(\omega_k)=1.
$
For each risky asset $m$, the discounted gain vector
$
\left(
\Delta S_m^*(\omega_1),\ldots,
\Delta S_m^*(\omega_N)
\right)^\top
$
belongs to $U_1$. Therefore,
$
\sum_{k=1}^{N}
q_k\Delta S_m^*(\omega_k)
=0.
$
Dividing by $\sum_{j=1}^{N}q_j>0$ yields
\[
\sum_{k=1}^{N}
Q(\omega_k)\Delta S_m^*(\omega_k)
=0,
\qquad
m=1,\ldots,M.
\]
Hence,
$
E_Q[\Delta S_m^*]=0,
$ for all $m=1,\ldots,M$,
and $Q$ is a strictly positive risk-neutral probability measure.
This completes the proof.
\end{proof}

\begin{remark}
	\label{re:M(Q)}
	Since financial markets are typically incomplete, the risk neutral probability measure is generally not unique.
	Accordingly, we consider all such measures and define
	\begin{equation}
	\label{eq:M(Q)}
		\mathcal{M}(Q)= \left\{ Q:\ Q(\omega)>0,\ \forall\ \omega\in\Omega,
		\quad
		\sum_{\omega\in\Omega}Q(\omega)=1,
		\quad
		E_Q[\Delta S_m^*]=0,\ 
		m=1,\ldots,M
		\right\}.
	\end{equation}
	The probability measure $Q$ in $\mathcal M(Q)$ is a risk neutral measure associated with the discounted asset prices.
	Theorem \ref{theo:fundamental} requires $Q$ to be strictly positive on every state, but does not require $Q$ to belong to $\mathcal P$ or to be equivalent to any particular probability measure in $\mathcal P$.
	Here, $\mathcal P$ describes the family of possible models, whereas $\mathcal M(Q)$ collects all strictly positive probability measures under which the discounted gains of the risky securities have zero expectation.
	This distinction will be important when we consider nonlinear valuations and superhedging prices under model uncertainty.
\end{remark}

Based on Theorem \ref{theo:fundamental}, the value of a self-financing portfolio can be characterized by a risk neutral probability measure $Q$.
Specifically,
$$
V_0=V_0^*
=h_0+\sum_{m=1}^{M}h_mE_Q[S_m^*(1)]
=E_Q\left[\frac{V_1}{S_0(1)}\right],
$$
where $Q\in\mathcal M(Q)$.
Since $V_1$ is the terminal value of a self-financing portfolio, the expectation above is the same for every $Q\in\mathcal M(Q)$.
However, in an incomplete market, a contingent claim that cannot be replicated by the traded securities may have different expected payoffs under different risk neutral probability measures.
Consequently, the family $\mathcal M(Q)$ gives rise to a range of possible valuations for such claims rather than a unique price.
This distinction provides the basis for studying nonlinear valuations and their relationship with superhedging prices in the subsequent sections.

\subsection{Asset pricing under model uncertainty with short sales prohibitions}

Short-selling of stocks is prohibited in some financial markets, and such restrictions may affect market stability and price formation.
For example, \citet{Battalio11} documented that, during short-sale bans, synthetic share prices for restricted stocks can become substantially lower than their observed market prices.
Short-sale prohibitions also restrict the trading strategies available to investors and may increase the cost of hedging contingent claims \citep{He20}.
In this study, we incorporate short-sale prohibitions by requiring that the holdings in risky securities are nonnegative, i.e., $\hat h_m\geq0$, $m=1,\ldots,M$, while no restriction is imposed on the position in the bond $h_0$.
Under these constraints, the set of attainable discounted gains is a convex cone rather than a linear space.
Consequently, the martingale condition in the unrestricted market is replaced by the corresponding supermartingale inequality.
The relationship between no-arbitrage and supermartingale measures under portfolio constraints has been studied in the literature, including \citet{Pulido14} and \citet{BayraktarZhou2017}.
We first present this characterization in the finite-state model, which will be used in the subsequent analysis of nonlinear valuations and superhedging prices.

To characterize the absence of arbitrage under model uncertainty in the presence of short sales prohibitions, we consider both individual risk neutral probability measures and families of probability measures, and investigate their relationship with the no-arbitrage condition, see Figure \ref{fig:arbitrage short}.
We first introduce the risk neutral probability measure under short sales prohibitions in Definition \ref{de:risk neutral short}, and establish the equivalence between its existence and the no-arbitrage condition.
We then consider whether the conditions imposed on an individual risk neutral probability measure can be relaxed when a family of probability measures $\mathcal{Q}$ is considered.
Instead of requiring every probability measure in $\mathcal{Q}$ to satisfy Definition \ref{de:risk neutral short}, we impose conditions on the lower or upper expectations of the discounted asset gains, together with a support condition on the family as a whole.
This leads to the weak and strong risk neutral nonlinear expectation conditions in Definitions \ref{de:risk neutral weak expectation} and \ref{de:risk neutral strong expectation}, respectively.
Based on these two definitions, we investigate the necessary and sufficient conditions for no-arbitrage under model uncertainty with short sales prohibitions.
In particular, the no-arbitrage condition implies the existence of a family satisfying the weak risk neutral nonlinear expectation condition.
We further show that the existence of a family satisfying the strong risk neutral nonlinear expectation condition ensures the absence of arbitrage.

\begin{figure}[htbp]
	\centering
	\begin{equation*}
			\begin{matrix}
					\text{No-arbitrage under model uncertainty} &\multirow{2}{*}{\(\Longleftrightarrow\)} &\text{Risk neutral probability measure} \\
					\text{{Short sales prohibitions}}& & {E_Q[\Delta S_m^*]\le 0,\ Q(\omega)>0}\\
					\Uparrow  &  & \Downarrow \\
					\text{Strong risk neutral nonlinear expectation} & \multirow{2}{*}{\(\Longrightarrow\)} & \text{Weak risk neutral nonlinear expectation}\\
					{\sup_{Q\in\mathcal{Q}}E_Q[\Delta S_m^*]\le 0,\ \sup_{Q\in\mathcal{Q}}Q(\omega)>0}  & & {\inf_{Q\in\mathcal{Q}}E_Q[\Delta S_m^*]\le 0,\ \sup_{Q\in\mathcal{Q}}Q(\omega)>0}
				\end{matrix}
		\end{equation*}
	\caption{Asset pricing with short sales prohibitions under model uncertainty}
	\label{fig:arbitrage short}
\end{figure}

We first define a probability measure under which the discounted prices of risky securities are supermartingales. In the presence of short-sale prohibitions, we refer to such a measure as a risk neutral probability measure.

\begin{definition}[\textbf{Risk neutral probability measure}]
	\label{de:risk neutral short}
	With short sales prohibitions, a probability measure $Q$ on $\Omega$ is said to be a risk neutral probability measure if it satisfies
	\begin{equation}
		\label{eq:risk neutral short}
		E_Q[\Delta S_m^*]\le 0,\quad m=1,\cdots,M,
	\end{equation}
	where $Q(\omega)>0$ for all $\omega\in\Omega$.
\end{definition}

Since $S_m^*(0)$ is deterministic, equation \eqref{eq:risk neutral short} is equivalent to
$$
S_m^*(0)\geq E_Q[S_m^*(1)],
\qquad m=1,\ldots,M.
$$
Thus, each discounted risky-asset price is a supermartingale under $Q$.
We next establish the relationship between the no-arbitrage condition and the existence of such a probability measure.

\begin{theorem}[\textbf{Fundamental theorem of asset pricing}]
	\label{theo:fundamental short}
	Under short sales prohibitions, no-arbitrage condition under model uncertainty holds if and only if there exists a risk neutral probability measure $Q$.
\end{theorem}
\begin{proof}

For the "if" direction, suppose that $Q$ satisfies Definition \ref{de:risk neutral short}.
For any admissible portfolio with $\hat h_m\geq0$, $m=1,\ldots,M$, we have
\[
\begin{aligned}
	E_Q[V_1^*]
	&=h_0+\sum_{m=1}^{M}\hat h_m E_Q[S_m^*(1)]\\
	&=V_0^*+\sum_{m=1}^{M}\hat h_m E_Q[\Delta S_m^*]\\
	&\leq V_0^*.
\end{aligned}
\]
Suppose that an arbitrage opportunity exists.
Then $V_0^*=0$, $V_1^*\geq0$ in every state, and
$
\sup_{P\in\mathcal P}E_P[V_1^*]>0.
$
The last condition implies that $V_1^*$ is strictly positive in at least one state.
Since $Q$ is strictly positive on $\Omega$, we have
$
E_Q[V_1^*]>0.
$
This contradicts $E_Q[V_1^*]\leq V_0^*=0$.
Hence, no-arbitrage holds.

For the converse, suppose that the no-arbitrage condition under model uncertainty holds. Define the set of attainable discounted gains under short sales prohibitions by
\begin{equation}
	U_1
	:=
	\left\{
	\left(
	G^*(\omega_1),\ldots,G^*(\omega_N)
	\right)^\top
	:
	G^*
	=
	\sum_{m=1}^{M}\hat h_m\Delta S_m^*,
	\quad
	\hat h\in\mathbb R_+^M
	\right\},
	\label{eq:U1-short-gain}
\end{equation}
and let
\begin{equation}
	U_2:=\mathbb R_+^N.
	\label{eq:U2-short-gain}
\end{equation}
Since the admissible risky-asset positions satisfy $\hat h_m\geq0$ for all $m=1,\ldots,M$, the set $U_1$ is the convex cone generated by the finite collection of discounted gain vectors
$
\left(
\Delta S_m^*(\omega_1),\ldots,
\Delta S_m^*(\omega_N)
\right)^\top,
\ m=1,\ldots,M.
$
Hence, $U_1$ is a finitely generated polyhedral cone and is therefore closed and convex.
By the same argument as in the converse proof of Theorem \ref{theo:fundamental}, the no-arbitrage condition, together with the support assumption
$
\sup_{P\in\mathcal P}P(\omega)>0,
\ \forall\omega\in\Omega,
$
implies that
\begin{equation}
	U_1\cap U_2=\{0\}.
	\label{eq:intersection-short-gain}
\end{equation}
Indeed, suppose that a nonzero vector $y$ belongs to $U_1\cap U_2$. Then there exists $\hat h\in\mathbb R_+^M$ such that
\[
y=\sum_{m=1}^{M}\hat h_m
\left(
\Delta S_m^*(\omega_1),\ldots,
\Delta S_m^*(\omega_N)
\right)^\top.
\]
By choosing the position in the bond as
$
h_0=-\sum_{m=1}^{M}\hat h_mS_m^*(0),
$
we obtain a portfolio with $V_0^*=0$ and $V_1^*(\omega_k)=y_k$, $k=1,\ldots,N$. Since $y\in\mathbb R_+^N$ and $y\neq0$, we have $V_1^*(\omega)\geq0$ for all $\omega\in\Omega$ and $V_1^*(\omega_{k_0})>0$ for at least one state $\omega_{k_0}$. By the support assumption on $\mathcal P$,
\[
\sup_{P\in\mathcal P}E_P[V_1^*]
\geq
V_1^*(\omega_{k_0})
\sup_{P\in\mathcal P}P(\omega_{k_0})
>0,
\]
which contradicts the no-arbitrage condition in Definition \ref{de:arbitrage uncertainty}.
To construct a strictly positive probability measure, we consider the compact convex set
\[
D:=
\left\{
x\in\mathbb R_+^N:
\sum_{k=1}^{N}x_k=1
\right\}.
\]
Since $D\subset U_2\setminus\{0\}$, equation \eqref{eq:intersection-short-gain} implies that $U_1\cap D=\varnothing$.
As $U_1$ is closed and convex and $D$ is compact and convex, by the finite-dimensional strong separating-hyperplane theorem, there exists a nonzero vector
$
q=(q_1,\ldots,q_N)^\top\in\mathbb R^N
$
such that
\[
\sup_{y\in U_1}q^\top y
<
\inf_{x\in D}q^\top x.
\]
Since $U_1$ is a cone containing the origin, we must have
\begin{equation}
	q^\top y\leq0,
	\qquad \forall\,y\in U_1.
	\label{eq:q-cone-short}
\end{equation}
Otherwise, if $q^\top y_0>0$ for some $y_0\in U_1$, then $\lambda y_0\in U_1$ for every $\lambda>0$, and consequently
\[
\sup_{y\in U_1}q^\top y
\geq
\lambda q^\top y_0
\longrightarrow+\infty,
\qquad \lambda\to+\infty,
\]
which contradicts the separating inequality.
Since $0\in U_1$, it follows that
$
\sup_{y\in U_1}q^\top y=0,
$
and hence
\[
\inf_{x\in D}q^\top x>0.
\]
For each $k=1,\ldots,N$, the standard unit vector $e_k$ belongs to $D$. Therefore,
$
q_k=q^\top e_k>0,
$
and thus $q\in\mathbb R_{++}^N$.
Define a probability measure $Q$ on $\Omega$ by
\begin{equation}
	Q(\omega_k)
	:=
	\frac{q_k}{\sum_{j=1}^{N}q_j},
	\qquad k=1,\ldots,N.
	\label{eq:Q-short-gain}
\end{equation}
Since $q_k>0$ for every $k$, we have
$
Q(\omega_k)>0,
\ k=1,\ldots,N,
$
and
$
\sum_{k=1}^{N}Q(\omega_k)=1.
$
Thus, $Q$ is a strictly positive probability measure.
For each risky security $m$, choose the admissible strategy satisfying
\[
\hat h_m=1
\qquad\text{and}\qquad
\hat h_j=0,\quad j\neq m.
\]
The corresponding discounted gain vector
$
\left(
\Delta S_m^*(\omega_1),\ldots,
\Delta S_m^*(\omega_N)
\right)^\top
$
belongs to $U_1$. It follows from \eqref{eq:q-cone-short} that
\[
\sum_{k=1}^{N}
q_k\Delta S_m^*(\omega_k)
\leq0.
\]
Dividing by $\sum_{j=1}^{N}q_j>0$ yields
\[
\sum_{k=1}^{N}
Q(\omega_k)\Delta S_m^*(\omega_k)
\leq0,
\qquad m=1,\ldots,M.
\]
Therefore,
\[
E_Q[\Delta S_m^*]\leq0,
\qquad m=1,\ldots,M.
\]
Hence, $Q$ is a risk neutral probability measure under short sales prohibitions, which completes the proof.
\end{proof}

\begin{remark}
	\label{re:M-(Q)}
	Similar to Remark \ref{re:M(Q)}, the risk neutral probability measure under short sales prohibitions need not be unique.
	We collect all such measures into the set $\tilde{\mathcal{M}}(Q)$, defined by
	$$
	\tilde{\mathcal{M}}(Q) =\{Q:\ Q(\omega)>0, \forall\ \omega\in\Omega,
	\quad
	\sum_{\omega\in\Omega}Q(\omega)=1,
	\quad
	E_Q[\Delta S_m^*]\le 0,\ m=1,\cdots,M\}.
	$$
	In the single-model case, where $\mathcal P=\{P\}$ and $P(\omega)>0$ for all $\omega\in\Omega$, the condition $Q(\omega)>0$ for all $\omega\in\Omega$ is equivalent to $Q\sim P$.
	In this case, Theorem \ref{theo:fundamental short} corresponds to the finite-state, single-period version of the fundamental theorem of asset pricing under short sales prohibitions studied by \citet{Pulido14}.
	Under model uncertainty, however, $\mathcal P$ is a family of probability measures, and the support condition only requires that each state has positive probability under at least one measure in $\mathcal P$.
	Therefore, we do not require $Q$ to be equivalent to every $P\in\mathcal P$, nor do we require $Q\in\mathcal P$.
\end{remark}

In the absence of short-selling constraints, market participants can freely take long and short portfolios, and the no-arbitrage condition is equivalent to the existence of a family of risk neutral probability measures $\mathcal{M}(Q)$ (see Theorem \ref{theo:fundamental} and Remark \ref{re:M(Q)}).
However, in markets with short sales prohibitions, trading strategies are restricted and risks cannot be hedged through short positions.
This further increases market incompleteness \citep{He20}, implying that no-arbitrage prices cannot be fully characterized by the family of risk neutral probability measures $\tilde{\mathcal{M}}(Q)$ (see Theorem \ref{theo:fundamental short} and Remark \ref{re:M-(Q)}).
To address this, we introduce a broader family of probability measures $\mathcal{Q}$ for further analysis.
This induces a corresponding family of expectations $\{E_Q[\cdot]: Q\in\mathcal{Q}\}$, based on which we construct the weak and strong risk neutral nonlinear expectations conditions.

\begin{definition}[\textbf{Weak risk neutral nonlinear expectation}]
	\label{de:risk neutral weak expectation}
	With short sales prohibitions, the family $\{E_Q[\cdot]:Q\in\mathcal Q\}$ is said to define a weak risk-neutral nonlinear expectation if
	\begin{equation}
		\label{eq:risk neutral weak expectation}
		\inf_{Q\in\mathcal{Q}} E_{Q}[\Delta S_m^*]\le 0, \quad m=1,\cdots,M,
	\end{equation}
	where $\sup_{Q\in\mathcal{Q}}Q(\omega)>0$ for all $\omega\in\Omega$.
\end{definition}

\begin{definition}[\textbf{Strong risk neutral nonlinear expectation}]
	\label{de:risk neutral strong expectation}
	With short sales prohibitions, the family $\{E_Q[\cdot]:Q\in\mathcal Q\}$ is said to define a strong risk-neutral nonlinear expectation if
	\begin{equation}
		\label{eq:risk neutral strong expectation}
		\sup_{Q\in\mathcal{Q}} E_{Q}[\Delta S_m^*]\le 0, \quad m=1,\cdots,M,
	\end{equation}
	where $\sup_{Q\in\mathcal{Q}}Q(\omega)>0$ for all $\omega\in\Omega$.
\end{definition}

\begin{remark}
	\label{re:strong&weak}
	Note that $\inf_{Q\in\mathcal{Q}} E_{Q}[\Delta S_m^*]\le\sup_{Q\in\mathcal{Q}} E_{Q}[\Delta S_m^*]$, so the strong risk neutral nonlinear expectation condition implies the weak one.
	Compared with Definition \ref{de:risk neutral short}, the strong condition allows probability measures that are not strictly positive on $\Omega$, provided that the family $\mathcal{Q}$ satisfies the support condition.
	The weak condition further relaxes the requirement on the expected discounted gains by considering the lower expectation instead of the upper expectation.
	If the family $\mathcal{Q}$ consists of a single probability measure $Q$, Definitions \ref{de:risk neutral weak expectation} and \ref{de:risk neutral strong expectation} coincide with Definition \ref{de:risk neutral short}.
\end{remark}

The weak and strong risk neutral nonlinear expectations provide necessary and sufficient conditions for no-arbitrage under model uncertainty with short sales prohibitions. We first establish the necessary condition for no-arbitrage.
\begin{theorem}[\textbf{Necessary condition for no-arbitrage under model uncertainty}]
	\label{theo:arbitrage and weak}
	With short sales prohibitions, no-arbitrage under model uncertainty guarantee the existence of a weak risk neutral nonlinear expectation.
\end{theorem}
\begin{proof}
	By Theorem \ref{theo:fundamental short}, the no-arbitrage condition implies the existence of a risk neutral probability measure $Q$ under short sales prohibitions such that $E_Q[\Delta S_m^*]\le 0$, where $Q(\omega)>0$ for all $\omega\in\Omega$.
	Including this measure $Q$ in the family $\mathcal{Q}$, we obtain
	$$
	\inf_{Q\in\mathcal{Q}} E_Q[\Delta S_m^*]\le E_Q[\Delta S_m^*]\le 0.
	$$
	Moreover, since $Q(\omega)>0$ for all $\omega\in\Omega$, it follows that $\sup_{Q\in\mathcal{Q}} Q(\omega)\ge Q(\omega)>0,\ \forall \omega\in\Omega$.
	Therefore, a weak risk neutral nonlinear expectation exists, which completes the proof.
\end{proof}

For the weak risk neutral nonlinear expectation, we have
\begin{equation}
	\label{eq:inf}
	\sum_{m=1}^{M} \inf_{Q\in\mathcal{Q}} E_Q[\Delta S_m^*]\le \inf_{Q\in\mathcal{Q}} E_Q[\sum_{m=1}^{M} \Delta S_m^*].
\end{equation}

The inequality \eqref{eq:inf} shows that the lower expectations of individual asset gains do not necessarily provide the same bound for the gain of a portfolio.
In particular, the probability measures that give nonpositive expected gains may differ from one security to another.
We give a simple example to illustrate this distinction.

\begin{example}
	Consider a market with two states $\Omega=\{\omega_1,\omega_2\}$ and two risky securities.
	Let
	$$
	\Delta S_1^*=(1,-1),
	\qquad
	\Delta S_2^*=(-1,2).
	$$
	Take $\mathcal Q=\{Q_1,Q_2\}$, where
	$$
	Q_1=(1/4,3/4),
	\qquad
	Q_2=(3/4,1/4).
	$$
	Both probability measures are strictly positive on $\Omega$.
	For the first security, we have
	$$
	\inf_{Q\in\mathcal Q}E_Q[\Delta S_1^*]
	=-\frac12,
	$$
	whereas for the second security,
	$$
	\inf_{Q\in\mathcal Q}E_Q[\Delta S_2^*]
	=-\frac14.
	$$
	Hence, the weak risk neutral nonlinear expectation condition is satisfied.
	However, the portfolio with $\hat h_1=\hat h_2=1$ has discounted gain
	$$
	G^*=\Delta S_1^*+\Delta S_2^*=(0,1).
	$$
	By choosing the position in the bond so that the initial portfolio value is zero, we obtain an arbitrage opportunity.
	Indeed, the discounted terminal gain is nonnegative in every state and strictly positive in $\omega_2$.
	The support condition on $\mathcal P$ then gives
	$
	\sup_{P\in\mathcal P}E_P[G^*]>0.
	$
	Therefore, the weak risk neutral nonlinear expectation condition alone does not guarantee no-arbitrage.
\end{example}

We next consider the strong risk neutral nonlinear expectation and establish a sufficient condition for no-arbitrage.
\begin{theorem}[\textbf{Sufficient condition for no-arbitrage under model uncertainty}]
	\label{theo:arbitrage and strong}
	With short sales prohibitions, the existence of a strong risk neutral nonlinear expectation guarantees no-arbitrage under model uncertainty.
\end{theorem}
\begin{proof}
		Suppose, to the contrary, that an arbitrage opportunity exists.
		Then
		$
		V_0^*=0,
		$
		$
		V_1^*(\omega)\geq0,\quad\forall\omega\in\Omega,
		$
		and
		$
		\sup_{P\in\mathcal P}E_P[V_1^*]>0.
		$
		The last condition implies that $V_1^*(\omega_0)>0$ for some $\omega_0\in\Omega$.
		By the strong risk neutral nonlinear expectation condition and $\hat h_m\geq0$, $m=1,\ldots,M$, we obtain
		$$
		V_0^*
		=h_0+\sum_{m=1}^{M}\hat h_m S_m^*(0)\geq h_0+\sum_{m=1}^{M}\hat h_m
		\sup_{Q\in\mathcal Q}E_Q[S_m^*(1)]
		\geq
		\sup_{Q\in\mathcal Q}E_Q[V_1^*].
		$$
		The first inequality follows from
		$
		\sup_{Q\in\mathcal Q}E_Q[\Delta S_m^*]\leq0,
		$
		and the second follows from the subadditivity of the upper expectation.
		
		Since $V_1^*\geq0$ and $V_1^*(\omega_0)>0$, the support condition on $\mathcal Q$ implies that
		$$
		\sup_{Q\in\mathcal Q}E_Q[V_1^*]\geq V_1^*(\omega_0)\sup_{Q\in\mathcal Q}Q(\omega_0)>0.
		$$
		Consequently, $V_0^*>0$, which contradicts $V_0^*=0$.
		Therefore, the no-arbitrage condition under model uncertainty holds.
\end{proof}

\begin{remark}
	\label{re:degenerate}
	If the family $\mathcal Q$ consists of a single probability measure $Q$, the weak and strong risk neutral nonlinear expectation conditions coincide.
	In this case, both Definitions \ref{de:risk neutral weak expectation} and \ref{de:risk neutral strong expectation} reduce to Definition \ref{de:risk neutral short}.
	Consequently, the corresponding no-arbitrage characterization is given by Theorem \ref{theo:fundamental short}.
\end{remark}

Based on Theorem \ref{theo:arbitrage and strong}, we further examine the valuation of a portfolio under the strong risk neutral nonlinear expectation.
For any admissible portfolio with $\hat h_m\geq0$, $m=1,\ldots,M$, we have
$$
\begin{aligned}
	V_0^*
	&=h_0+\sum_{m=1}^{M}\hat h_m S_m^*(0)\geq h_0+\sum_{m=1}^{M}\hat h_m
	\sup_{Q\in\mathcal Q}E_Q[S_m^*(1)]\\
	&\geq \sup_{Q\in\mathcal Q}
	E_Q\left[\frac{V_1}{S_0(1)}\right].
\end{aligned}
$$
The first inequality follows from the strong risk neutral nonlinear expectation condition and the nonnegativity of the risky-asset positions.
The second inequality follows from the subadditivity of the upper expectation.
This relation shows that the upper expectation over $\mathcal Q$ provides a lower bound for the initial value of any admissible portfolio.
Since $S_0(0)=1$, we have $V_0=V_0^*$.
We therefore use
$
\sup_{Q\in\mathcal Q}E_Q[V_1/S_0(1)]
$
as the risk neutral nonlinear valuation under model uncertainty with short-sale prohibitions.

\begin{remark}
	\label{re:sublinear}
	The supremum of linear expectations
	$
	\sup_{Q\in\mathcal Q}E_Q[V_1/S_0(1)]
	$
	defines a sublinear expectation, denoted by
	$
	\mathbb E[V_1/S_0(1)]
	$
	(see \citet{Peng2019}).
	In the discrete-time, finite-state setting, this sublinear expectation admits an explicit representation through repeated summation; see \citet{YZ26} for details.
\end{remark}

\section{Multi-period asset pricing}
\label{sec:multi}

In this section, we extend the pricing model from the single-period setting to the multi-period case.
We consider a $T$-period discrete-time model on a finite event tree with terminal state space $\Omega$ and filtration $\mathbb F=(\mathcal F_t)_{t=0}^{T}$. 
Each node $n$ at time $t$, $t=0,\ldots,T-1$, has a finite set of successor nodes $\operatorname{succ}(n)$.
A local probability kernel $q_t(\cdot\mid n)$ assigns strictly positive probabilities to these successors, and the product of the local kernels along a path defines a global probability measure on $\Omega$. The portfolio consists of $M$ risky securities $S_m(t),\ m=1,2,\cdots,M$ and a bond $S_0(t)$, where $t=0,1,\cdots,T$. We take $S_0(0)=1$ and
$$
S_0(t)=(1+r_1)(1+r_2)\cdots(1+r_t),\quad t=1,2,\cdots,T,
$$
where $r_t$ is the interest rate over the period $(t-1,t)$, $t=1,\ldots,T$, is $\mathcal{F}_{t-1}$-measurable, and satisfies $1+r_t>0$.  
A trading strategy is a vector stochastic process $h(t)=(h_0(t), h_1(t), h_2(t),\cdots ,h_M(t))$, $t=1,2,\cdots,T$, where $h_m(t)$ denotes the number of units of security $m$ held over the period $(t-1, t)$, and is $\mathcal{F}_{t-1}$-measurable.
The discounted price process is defined by
$S_{m}^{*}(t)=S_m(t)/S_0(t)$,
and we denote $\Delta S_{m}^{*}(t)=S_{m}^{*}(t)-S_{m}^{*}(t-1)$, for $t=1,\cdots,T,\ m=1,2,\cdots,M$. The discounted value process $V^{*}(t)$ and discounted gain process $G^{*}(t)$ are given by
\begin{equation}
	\label{eq:value discount multi}
	V^{*}(t)=h_0(t)+\sum_{m=1}^{M}h_m(t)S^{*}_m(t),\quad t=1,2,\cdots,T.
\end{equation}
\begin{equation}
	\label{eq:gain discount multi}
	G^{*}(t)=\sum_{m=1}^{M} \sum_{u=1}^{t} h_m(u)\Delta S^{*}_m(u),\quad t=1,2,\cdots, T.
\end{equation}
The initial discounted wealth is
$
V^*(0)=h_0(1)+\sum_{m=1}^M h_m(1)S_m^*(0).
$
Let $t^{+}$ denote the instant immediately after rebalancing at time $t$. When the portfolio is updated from $h(t)$ to $h(t+1)$, the discounted portfolio value at $t^{+}$ is
\begin{equation}
	\label{eq:value t+}
	V^*(t^{+})=h_0(t+1)+\sum_{m=1}^{M}h_m(t+1)S^*_m(t).
\end{equation}
We adopt the self-financing condition, meaning that any change in the portfolio is financed by internal reallocation of wealth. Hence $V(t)=V(t^{+})$, which is equivalent to
\begin{equation}
	\label{eq:self finance}
	[h_0(t+1)-h_0(t)]+\sum_{m=1}^{M}[h_m(t+1)-h_m(t)]S^*_m(t)=0, \quad t=1,\ldots,T-1.
\end{equation}
By equation \eqref{eq:self finance}, a trading strategy $h$ is self-financing if and only if $V^{*}(t)=V^{*}(0)+G^{*}(t)$ holds for all $t=1,2,\cdots,T$.

\subsection{Pricing under model uncertainty}

We extend the fundamental theorem of asset pricing to the multi-period setting.
We first introduce the definition of arbitrage under model uncertainty.

\begin{definition}[\textbf{Multi-period Arbitrage under model uncertainty}]
	\label{de:arbitrage multi}
	A trading strategy $h$ is called an arbitrage under model uncertainty if the discounted value of the portfolio satisfies:
	
	(i) $\ V^*(0)=0$;
	
	(ii) $\ V^*(T; \omega)\ge 0,\ \forall \omega\in\Omega$ and $\sup_{P\in\mathcal{P}}E_P[V^*(T)]>0$, where $E_P[\cdot]$ is the expectation under actual probability measure $P\in\mathcal{P}$;
	
	(iii) $h$ is self-financing.
	
\end{definition}

\begin{remark}
	\label{re:arbitrage multi}
	By the self-financing condition, we have
	$V^*(T)=V^*(0)+G^*(T)$.
	Thus Definition \ref{de:arbitrage multi} is equivalent to
	(i) $V^*(0)=0$;
	(ii) $G^*(T;\omega)\geq 0$ for all $\omega\in\Omega$ and
	$\sup_{P\in\mathcal P}E_P[G^*(T)]>0$.
	If the multi-period market satisfies no-arbitrage, then each one-period market at every node of the event tree is also arbitrage-free.
	In particular, when $\mathcal P$ consists of a single full-support probability measure, Definition \ref{de:arbitrage multi} reduces to the classical notion of arbitrage in \citet{Kwok08}.
\end{remark}

\begin{remark}
	For a fixed probability measure $Q$, the conditional expectation
	$E_Q[\cdot\mid\mathcal F_t]$ satisfies the usual tower property.
	On a finite event tree, a collection of strictly positive local probability kernels
	$q_t(\cdot\mid n)$ determines a global probability measure by multiplying the
	transition probabilities along each path.
	For a family of probability measures, we assume rectangularity, or stability
	under nodewise pasting, whenever lower or upper expectations are evaluated
	recursively.
	This condition allows conditional probability kernels selected at different
	nodes to be combined while keeping the resulting measure in the same family.
	It also gives the tower property for the corresponding nonlinear expectations.
\end{remark}

We now characterize the no-arbitrage condition in terms of a risk-neutral probability measure.
Following \citet{Kwok08}, a probability measure $Q$ on $\Omega$ is called a risk-neutral probability measure if it satisfies
\begin{equation}
	\label{eq:martingale}
	E_Q[S_m^*(u)\mid\mathcal F_t]=S_m^*(t),
	\quad 0\leq t\leq u\leq T,\quad m=1,\ldots,M.
\end{equation}
We require $Q(\omega)>0$ for all $\omega\in\Omega$.
The fundamental theorem of asset pricing in the multi-period model is stated as follows.

\begin{theorem}[\textbf{Fundamental theorem of asset pricing under model uncertainty}]
	\label{theo:fundamental multi}
	In the multi-period model, the no-arbitrage condition under model uncertainty holds if and only if there exists a risk neutral probability measure $Q$.
\end{theorem}
\begin{proof}

Suppose first that a strictly positive martingale measure $Q$ exists.
For a predictable self-financing strategy $h$, we have
\[
	E_Q[G^*(T)]
	=\sum_{u=1}^{T}\sum_{m=1}^{M}
	E_Q[h_m(u)\Delta S_m^*(u)]
	=\sum_{u=1}^{T}\sum_{m=1}^{M}
	E_Q\!\left[
	h_m(u)E_Q[\Delta S_m^*(u)\mid\mathcal F_{u-1}]
	\right]
	=0.
\]
Therefore, $E_Q[V^*(T)]=V^*(0)$.
By the support assumption on $\mathcal P$, an arbitrage would satisfy
$V^*(T;\omega)>0$ for some $\omega\in\Omega$.
Since $Q(\omega)>0$ for every $\omega\in\Omega$, this gives
$E_Q[V^*(T)]>0$, contradicting $V^*(0)=0$.

Conversely, assume that the multi-period market satisfies no-arbitrage.
At any node $n$ at time $t$, a one-period arbitrage in the market from $n$ to its successor nodes could be extended to a self-financing strategy in the whole market.
Indeed, one may hold zero positions before reaching $n$, take the corresponding arbitrage positions at $n$, and hold the resulting wealth in the bond after time $t+1$.
The resulting terminal value would be nonnegative in every state and strictly positive in at least one terminal state.
By the support assumption on $\mathcal P$, this would give a multi-period arbitrage.
Hence every nodewise one-period market is arbitrage-free. The support assumption on $\mathcal P$ also gives the corresponding support condition on the successor nodes of each node.
By Theorem \ref{theo:fundamental}, each node admits a strictly positive local martingale kernel $q_t(\cdot\mid n)$ satisfying
\[
\sum_{n'\in\operatorname{succ}(n)}
q_t(n'\mid n)S_m^*(t+1,n')
=S_m^*(t,n),
\quad m=1,\ldots,M.
\]
Multiplying these kernels along each path defines a strictly positive global probability measure $Q$.
By construction,
\[
E_Q[S_m^*(t+1)\mid\mathcal F_t]=S_m^*(t),
\quad t=0,\ldots,T-1.
\]
The tower property then gives equation \eqref{eq:martingale} for all $0\leq t\leq u\leq T$.
Thus $Q$ is a multi-period risk-neutral probability measure, which completes the proof.
\end{proof}

Since the discounted asset prices are $Q$-martingales and the trading strategy is self-financing, the discounted portfolio value process $V^*$ is also a $Q$-martingale.
Consequently, for $0\leq t\leq T$, we have
\begin{equation}
	\label{eq:risk neutral valuation multi}
	V(t)
	=S_0(t)V^*(t)
	=S_0(t)E_Q[V^*(T)\mid\mathcal F_t]
	=S_0(t)E_Q\left[
	\frac{V(T)}{S_0(T)}
	\middle|\mathcal F_t
	\right].
\end{equation}
The martingale measure need not be unique in an incomplete market, even when the physical probability family $\mathcal P$ consists of a single full-support probability measure.
We collect all strictly positive martingale measures in the set
\begin{align}
	\label{eq:M'(Q)}
	\mathcal{M}'(Q)
	= \Biggl\{Q:\ &
	Q(\omega)>0,\quad \forall\,\omega\in\Omega,
	\quad
	\sum_{\omega\in\Omega}Q(\omega)=1,
	\notag\\
	& E_Q\!\left[S_m^*(u)\mid\mathcal{F}_t\right]
	= S_m^*(t),\quad
	0\le t\le u\le T,\quad
	m=1,\ldots,M
	\Biggr\}.
\end{align}
For the self-financing portfolio considered above, all measures in $\mathcal M'(Q)$ give the same value.
The family $\mathcal M'(Q)$ also provides the basis for pricing contingent claims that are not replicable, which we study in Section \ref{sec:hedging}.

\subsection{Pricing under model uncertainty with short sales prohibitions }

In this section, we consider a multi-period securities model with short sales prohibitions, that is, the trading strategy satisfies $h_m(t)\geq0$ for $m=1,\ldots,M$ and $t=1,\ldots,T$.
We examine the relationship between the no-arbitrage condition and several risk neutral probability measure conditions under model uncertainty.
We first establish the fundamental theorem of asset pricing under short sales prohibitions and then introduce the weak and strong risk neutral nonlinear expectation conditions, as illustrated in Figure \ref{fig:arbitrage short multi}.

\begin{figure}[htbp]
	\centering
	\begin{equation*}
		\begin{matrix}
			\text{No-arbitrage under model uncertainty} &\multirow{2}{*}{\(\Longleftrightarrow\)} &\text{Risk neutral probability measure} \\
			\text{with short sales prohibitions}& & E_Q[S_m^*(u)\mid\mathcal{F}_t]\le S_m^*(t),\ Q(\omega)>0\\
			\Uparrow  &  & \Downarrow \\
			\text{Strong risk neutral nonlinear expectation} & \multirow{3}{*}{\(\Longrightarrow\)} & \text{Weak risk neutral nonlinear expectation}\\
			\sup_{Q\in\mathcal{Q}}E_Q[S_m^*(u)\mid\mathcal{F}_t]\le S_m^*(t),  & & \inf_{Q\in\mathcal{Q}}E_Q[S_m^*(u)\mid\mathcal{F}_t]\le S_m^*(t),\\
			\sup_{Q\in\mathcal{Q}}Q(\omega)>0 & & \sup_{Q\in\mathcal{Q}}Q(\omega)>0
		\end{matrix}
	\end{equation*}
	\caption{Asset pricing under model uncertainty with short sales prohibitions}
	\label{fig:arbitrage short multi}
\end{figure}

\begin{definition}[\textbf{Risk neutral probability measure}]
	\label{de:risk neutral short multi}
	In multi-period, a probability measure $Q$ is called a risk neutral probability measure (or a supermartingale measure) with short sales prohibitions if it satisfies
	\begin{equation}
		\label{eq:risk neutral short multi}
		E_Q[S_m^*(u)\mid\mathcal{F}_{t}]\le S_m^*(t),\quad 0\le t\le u\le T,
	\end{equation}
	where $Q(\omega)>0$ for all $\omega\in\Omega$.
\end{definition}

\begin{theorem}[\textbf{Fundamental theorem of asset pricing under model uncertainty}]
	\label{theo:fundamental short multi}
	In multi-period, the no-arbitrage condition under model uncertainty holds if and only if there exists a risk neutral probability measure $Q$ with short sales prohibitions.
\end{theorem}

\begin{proof}
Suppose first that a strictly positive supermartingale measure $Q$ exists.
Since the risky positions are predictable and nonnegative, we have
\[
E_Q[G^*(T)]
=\sum_{u=1}^{T}\sum_{m=1}^{M}
E_Q[h_m(u)\Delta S_m^*(u)]
=\sum_{u=1}^{T}\sum_{m=1}^{M}
E_Q\!\left[
h_m(u)E_Q[\Delta S_m^*(u)\mid\mathcal F_{u-1}]
\right]\le0.
\]
Hence, by the self-financing condition,
$
E_Q[V^*(T)]\le V^*(0).
$
If an arbitrage opportunity exists, then $V^*(0)=0$, $V^*(T)\geq0$ in every state, and $V^*(T)>0$ in at least one state.
Since $Q(\omega)>0$ for all $\omega\in\Omega$, this implies
$
E_Q[V^*(T)]>0,
$
which contradicts $E_Q[V^*(T)]\leq V^*(0)=0$.
Therefore, the no-arbitrage condition holds.

Conversely, assume that the multi-period market satisfies no-arbitrage.
At any node $n$ at time $t$, a one-period arbitrage among its successor nodes would give rise to a multi-period arbitrage.
Indeed, one may hold zero positions before reaching $n$, take the corresponding admissible positions at $n$, and hold the resulting wealth in the bond after time $t+1$.
The resulting terminal value would be nonnegative in every state and strictly positive in at least one terminal state.
The support assumption on $\mathcal P$ then contradicts the no-arbitrage condition.
Hence every nodewise one-period market is arbitrage-free.
The support assumption on $\mathcal P$ also gives the corresponding support condition on the successor nodes of each node.
By Theorem \ref{theo:fundamental short}, each node admits a strictly positive local probability kernel $q_t(\cdot\mid n)$ satisfying
\[
\sum_{n'\in\operatorname{succ}(n)}
q_t(n'\mid n)S_m^*(t+1,n')
\le S_m^*(t,n),
\quad m=1,\ldots,M.
\]
Multiplying these kernels along each path defines a strictly positive global probability measure $Q$.
By construction,
\[
E_Q[S_m^*(t+1)\mid\mathcal F_t]
\le S_m^*(t),
\quad t=0,\ldots,T-1.
\]
The tower property then gives
\[
E_Q[S_m^*(u)\mid\mathcal F_t]
\le S_m^*(t),
\quad 0\le t\le u\le T.
\]
Thus $Q$ is a risk neutral probability measure under short sales prohibitions, which completes the proof.
\end{proof}

In an incomplete multi-period market, the strictly positive supermartingale measure is generally not unique.
We collect all such measures into the set:
\begin{align}
	\label{eq:M~'Q}
	\tilde{\mathcal{M}}'(Q)
	= \Biggl\{Q:\ &
	Q(\omega)>0,\ \forall\,\omega\in\Omega,
	\quad
	\sum_{\omega\in\Omega}Q(\omega)=1,
	\notag\\
	& E_Q\!\left[S_m^*(u)\mid\mathcal{F}_t\right]
	\le S_m^*(t),\quad
	0\le t\le u\le T,\quad
	m=1,\ldots,M
	\Biggr\}.
\end{align}
The set $\tilde{\mathcal M}'(Q)$ consists of strictly positive supermartingale measures.
We now consider a broader family of probability measures $\mathcal Q$ on $\Omega$.
We do not require each measure in $\mathcal Q$ to satisfy the supermartingale condition individually or to assign positive probability to every terminal state.
To ensure that the conditional expectations are well defined at every nonterminal node, we require each $Q\in\mathcal Q$ to assign positive probability to every such node.
Moreover, the family satisfies $\sup_{Q\in\mathcal Q}Q(\omega)>0$ for all $\omega\in\Omega$.
We introduce the following weak and strong risk neutral nonlinear expectation conditions in terms of the lower and upper conditional expectations over $\mathcal Q$.

\begin{definition}[\textbf{Weak risk neutral nonlinear expectation}]
	\label{de:risk neutral weak expectation multi}
	In multi-period, $E_Q[\cdot], Q\in\mathcal{Q}$ is said to be a weak risk neutral nonlinear expectation with short sales prohibitions if it satisfies
	\begin{equation}
		\label{eq:risk neutral weak multi}
		\inf_{Q\in\mathcal{Q}}E_Q[S_m^*(u)\mid\mathcal{F}_t]\le S_m^*(t),\quad 0\le t\le u\le T,
	\end{equation}
	where $\sup_{Q\in\mathcal{Q}}Q(\omega)>0$ for all $\omega\in\Omega$.
\end{definition}

\begin{definition}[\textbf{Strong risk neutral nonlinear expectation}]
	\label{de:risk neutral strong expectation multi}
	In multi-period, $E_Q[\cdot], Q\in\mathcal{Q}$ is said to be a strong risk neutral nonlinear expectation with short sales prohibitions if it satisfies
	\begin{equation}
		\label{eq:risk neutral strong multi}
		\sup_{Q\in\mathcal{Q}}E_Q[S_m^*(u)\mid\mathcal{F}_t]\le S_m^*(t),\quad 0\le t\le u\le T,
	\end{equation}
	where $\sup_{Q\in\mathcal{Q}}Q(\omega)>0$ for all $\omega\in\Omega$.
\end{definition}

\begin{corollary}[\textbf{Necessary condition for no-arbitrage under model uncertainty}]
	\label{theo:arbitrage and weak multi}
	In the multi-period model, no-arbitrage under model uncertainty guarantees the existence of a weak risk neutral nonlinear expectation with short sales prohibitions.
\end{corollary}
\begin{proof}
By Theorem \ref{theo:fundamental short multi}, there exists a strictly positive supermartingale measure $Q^0$.
Take $\mathcal Q=\{Q^0\}$.
Since
\[
E_{Q^0}[S_m^*(u)\mid\mathcal F_t]\le S_m^*(t),
\quad 0\le t\le u\le T,\quad m=1,\ldots,M,
\]
both the weak and strong risk neutral nonlinear expectation conditions are satisfied.
Moreover, $Q^0(\omega)>0$ for all $\omega\in\Omega$, so the support condition also holds.
This completes the proof.
\end{proof}

\begin{theorem}[\textbf{Sufficient condition for no-arbitrage under model uncertainty}]
	\label{theo:arbitrage and strong multi}
	In the multi-period model, the existence of a strong risk neutral nonlinear expectation with short sales prohibitions guarantees no-arbitrage under model uncertainty.
\end{theorem}

\begin{proof}
	Suppose, to the contrary, that an arbitrage opportunity exists.
	Then $V^*(0)=0$, $V^*(T)\geq0$ in every state, and $V^*(T)>0$ in at least one terminal state $\omega_0$.
	By the strong risk neutral nonlinear expectation condition, for every $Q\in\mathcal Q$,
	\[
	E_Q[\Delta S_m^*(u)\mid\mathcal F_{u-1}]
	\leq0,
	\quad u=1,\ldots,T,\quad m=1,\ldots,M.
	\]
	Since the risky positions are predictable and nonnegative, we obtain
	\[
		E_Q[G^*(T)]
		=\sum_{u=1}^{T}\sum_{m=1}^{M}
		E_Q\!\left[
		h_m(u)E_Q[\Delta S_m^*(u)\mid\mathcal F_{u-1}]
		\right]
		\leq0.
	\]
	Consequently, by the self-financing condition,
	\[
	E_Q[V^*(T)]\leq V^*(0)=0,
	\qquad \forall Q\in\mathcal Q.
	\]
	On the other hand, the support condition
	$
	\sup_{Q\in\mathcal Q}Q(\omega_0)>0
	$
	implies that there exists $Q_{\omega_0}\in\mathcal Q$ such that
	$Q_{\omega_0}(\omega_0)>0$.
	Since $V^*(T)\geq0$ in every state and
	$V^*(T;\omega_0)>0$, we have
	\[
	E_{Q_{\omega_0}}[V^*(T)]>0,
	\]
	which contradicts
	$
	E_{Q_{\omega_0}}[V^*(T)]\leq0.
	$
	Therefore, the no-arbitrage condition holds.
\end{proof}

\begin{remark}
	If the family $\mathcal Q$ consists of a single probability measure $Q$, the weak and strong risk neutral nonlinear expectation conditions coincide with Definition \ref{de:risk neutral short multi}.
	In this case, the necessary and sufficient conditions for no-arbitrage are given by Theorem \ref{theo:fundamental short multi}.
\end{remark}

For the dynamic valuation below, we further assume that the family $\mathcal Q$ is rectangular.
By the strong risk neutral nonlinear expectation condition, the discounted value process of an admissible self-financing portfolio is a $Q$-supermartingale for every $Q\in\mathcal Q$.
Consequently, for $0\leq t\leq T$, we have
\[
\begin{aligned}
	V(t)
	&=S_0(t)V^*(t)
	\geq S_0(t)\sup_{Q\in\mathcal Q}
	E_Q[V^*(T)\mid\mathcal F_t]\\
	&=S_0(t)\sup_{Q\in\mathcal Q}
	E_Q\left[
	\frac{V(T)}{S_0(T)}
	\middle|\mathcal F_t
	\right].
\end{aligned}
\]
The rectangularity of $\mathcal Q$ ensures the time consistency of the corresponding upper expectation.
This quantity gives a lower bound for the current value of any admissible portfolio with the prescribed terminal value.
It should be distinguished from the exact constrained superhedging price, which is determined by the supremum over the complete dual family of supermartingale measures associated with the short-sale constraint.
The two quantities coincide when the supremum over the selected family $\mathcal Q$ equals that over the complete dual family for the claim under consideration.
This distinction will be made explicit in the superhedging duality developed in Section \ref{sec:hedging}.

\section{Multi-period hedging strategy}
\label{sec:hedging}

Hedging theorem serves as an immediate corollary of the fundamental theorem of asset pricing. Consequently, we first consider the exact hedging of a contingent claim $F$ and then investigate the superhedging problem when exact replication is not available.
For a predictable trading strategy $H=(h_m(u))$, where $m=1,\ldots,M$ and $u=1,\ldots,T$, the total discounted gain of the risky securities at time $t$ is defined by
\[
H\cdot S^*(t)
=
\sum_{m=1}^{M}\sum_{u=1}^{t}
h_m(u)\Delta S_m^*(u),
\quad t=1,\ldots,T,
\]
where
$\Delta S_m^*(u)=S_m^*(u)-S_m^*(u-1)$.
The holding in the bond is determined by the self-financing condition.
Under model uncertainty, the hedging conditions are understood with respect to the probability family $\mathcal P$ introduced above.

\subsection{Hedging and superhedging under model uncertainty}

We first define exact hedging and derive the corresponding risk neutral price.
We then consider superhedging, which requires the terminal portfolio value to dominate the contingent claim.

\begin{definition}[\textbf{Hedging strategy under model uncertainty}]
	\label{de:hedging}
	A contingent claim $F$ is perfectly hedged by a portfolio if there exist initial investment $x$ and trading strategy $H$ such that
	\begin{equation}
		\label{eq:hedge}
		x+H\cdot S^*(T)=F^*,\quad \mathcal{P}-q.s.,
	\end{equation}
	holds, where $F^*=F/S_0(T)$ is the discounted payoff of contingent claim,  $\mathcal{P}$ is the actual probability set,
	and a property holds $\mathcal{P}$-q.s. ($\mathcal{P}$-quasi-surely) if it holds outside a polar set $A$ which satisfies $\sup_{P\in\mathcal{P}} P(A)=0$.
\end{definition}

Definition \ref{de:hedging} implies that we need to determine the initial investment $x$ and trading strategy $H$ for hedging contingent claim $F$. Trading strategy can be obtained from equation \eqref{eq:hedge}.
Therefore, our main objective is to determine the initial investment $x$.
Based on Theorem \ref{theo:fundamental multi} and Remark \ref{re:M(Q)}, we can derive the hedging price under model uncertainty.
\begin{theorem}[\textbf{Hedging theorem under model uncertainty}]
	\label{theo:hedge}
	In multi-period, we assume that there is no-arbitrage under model uncertainty. If $F^*$ is replicable, then its hedging price is given by
	\begin{align}
		\label{eq:hedge price}
		\Pi(F):=&\{x\in\mathbb{R}:\exists\ H\ \text{such that}\ x+H\cdot S^*(T)= F^*,\ \mathcal{P}-q.s.\} \\
		=&E_Q[F^*], \quad \forall Q\in\mathcal{M}'(Q), \notag
	\end{align}
	where $\mathcal{M}'(Q)$ denotes the set of risk neutral probability measures defined in \eqref{eq:M'(Q)}. 
\end{theorem}

\begin{proof}
	By the support assumption
	$
	\sup_{P\in\mathcal P}P(\omega)>0,
	\ \forall\omega\in\Omega,
	$
	the only $\mathcal P$-polar set is the empty set.
	Hence, equation \eqref{eq:hedge} holds statewise and therefore $Q$-almost surely for every probability measure $Q$ on $\Omega$.
	Taking expectations under $Q\in\mathcal M'(Q)$, we obtain
	\[
	x+E_Q[H\cdot S^*(T)]=E_Q[F^*].
	\]
	Since $H$ is predictable and the discounted asset prices are $Q$-martingales,
	\[
	\begin{aligned}
		E_Q[H\cdot S^*(T)]
		&=\sum_{u=1}^{T}\sum_{m=1}^{M}
		E_Q[h_m(u)\Delta S_m^*(u)]\\
		&=\sum_{u=1}^{T}\sum_{m=1}^{M}
		E_Q\!\left[
		h_m(u)E_Q[\Delta S_m^*(u)\mid\mathcal F_{u-1}]
		\right]=0.
	\end{aligned}
	\]
	Consequently,
	\[
	x=E_Q[F^*],
	\qquad \forall Q\in\mathcal M'(Q).
	\]
	Since $\mathcal M'(Q)$ is nonempty by Theorem \ref{theo:fundamental multi}, the initial investment $x$ is uniquely determined.
	This completes the proof.
\end{proof}

\begin{remark}
	\label{re:hedge}
	Theorem~\ref{theo:hedge} establishes the risk-neutral valuation of a replicable contingent claim under model uncertainty.
	In the present finite-state framework, the support condition
	$
	\sup_{P\in\mathcal P}P(\omega)>0,\ \forall \omega\in\Omega,
	$
	implies that the only $\mathcal P$-polar set is the empty set.
	Thus, $\mathcal P$-quasi-sure equality is equivalent to statewise equality, and the pricing identity holds for every $Q\in\mathcal M'(Q)$ without requiring an additional absolute-continuity or equivalence relation between $Q$ and the measures in $\mathcal P$.
	
	This differs from the framework of \citet{Bouchard15}, where the dual martingale measures are linked to the family of physical probability measures through absolute-continuity conditions, with the associated polar sets playing a role in the duality results.
\end{remark}

Exact hedging requires the contingent claim to be replicable.
For a general contingent claim, we therefore consider superhedging strategies and characterize the corresponding price.

\begin{definition}[\textbf{Superhedging strategy under model uncertainty}]
	\label{de:superhedging}
	A contingent claim $F$ is superhedged by a portfolio if there exist initial investment $x$ and trading strategy $H$ such that
	\begin{equation}
		\label{eq:suphedge}
		x+ H\cdot S^*(T)\ge F^*,\quad \mathcal{P}-q.s.,
	\end{equation}
	holds, where $F^*=F/S_0(T)$ is the discounted payoff of the contingent claim and $\mathcal P$ is the actual probability set.
\end{definition}

\begin{theorem}[\textbf{Superhedging theorem under model uncertainty}]
	\label{theo:suphedge}
	In multi-period, we assume that there is no-arbitrage under model uncertainty. If $F^*$ is super-replicable, then its superhedging price is given by
	\begin{align}
		\label{eq:suphedge price}
		\Pi^+(F):=&\min \{x\in\mathbb{R}:\exists\ H\ \text{such that}\ x+H\cdot S^*(T)\ge  F^*,\ \mathcal{P}-q.s.\} \\
		=&\sup_{Q\in\mathcal{M}'(Q)}E_Q[F^*], \notag
	\end{align}
	where $\mathcal{M}'(Q)$ denotes the set of risk neutral probability measures defined in \eqref{eq:M'(Q)}.
\end{theorem}

\begin{proof}
	For every $Q\in\mathcal M'(Q)$, the martingale property of the discounted asset prices and the predictability of $H$ imply
	\[
	E_Q[H\cdot S^*(T)]
	=
	\sum_{u=1}^{T}\sum_{m=1}^{M}
	E_Q[h_m(u)\Delta S_m^*(u)]
	=0.
	\]
	Hence every superhedge satisfies
	\[
	x\geq E_Q[F^*],
	\qquad \forall Q\in\mathcal M'(Q),
	\]
	and therefore
	\[
	\Pi^+(F)\geq
	\sup_{Q\in\mathcal M'(Q)}E_Q[F^*].
	\]
	
	For the reverse inequality, we write the superhedging problem as a finite-dimensional linear program.
	Let $\mathcal N_t$ denote the set of nodes at time $t$, and let $\Omega(n)$ denote the set of terminal states whose paths pass through node $n$.
	For $\omega\in\Omega$, write $n_t(\omega)$ for the node visited by the path $\omega$ at time $t$.
	The primal problem is
	\[
	\begin{aligned}
		\min_{x,\{h_m(t,n)\}} \quad &x\\
		\text{subject to}\quad
		&x+\sum_{t=1}^{T}\sum_{m=1}^{M}
		h_m(t,n_{t-1}(\omega))
		\Delta S_m^*(t,\omega)
		\geq F^*(\omega),
		\quad \forall\omega\in\Omega.
	\end{aligned}
	\]
	The dual variables are nonnegative weights $q_\omega$ assigned to the terminal states.
	The dual constraint corresponding to $x$ is
	\[
	\sum_{\omega\in\Omega}q_\omega=1.
	\]
	Since the risky-asset positions are unrestricted, the dual constraints corresponding to $h_m(t,n)$ are
	\[
	\sum_{\omega\in\Omega(n)}
	q_\omega\Delta S_m^*(t+1,\omega)=0,
	\quad
	n\in\mathcal N_t,\quad
	t=0,\ldots,T-1,\quad
	m=1,\ldots,M.
	\]
	These are the martingale flow constraints on the event tree.
	Denote by $\overline{\mathcal M}'(Q)$ the set of probability measures on $\Omega$ satisfying these constraints, where zero probabilities at terminal states are allowed.
	For strictly positive probability measures, the flow constraints are equivalent to the martingale conditions in \eqref{eq:M'(Q)}.
	Thus,
	$
	\mathcal M'(Q)\subseteq\overline{\mathcal M}'(Q).
	$
	The dual objective is
	\[
	\sum_{\omega\in\Omega}q_\omega F^*(\omega)
	=E_Q[F^*].
	\]
	Since the primal problem is feasible and bounded below, strong linear-programming duality yields
	\[
	\Pi^+(F)
	=
	\max_{Q\in\overline{\mathcal M}'(Q)}
	E_Q[F^*].
	\]
	By Theorem \ref{theo:fundamental multi}, there exists a strictly positive martingale measure
	$Q^0\in\mathcal M'(Q)$.
	For any $Q\in\overline{\mathcal M}'(Q)$ and $\varepsilon\in(0,1)$, define
	$
	Q^\varepsilon
	:=(1-\varepsilon)Q+\varepsilon Q^0.
	$
	Since the martingale flow constraints are linear in the terminal-state probabilities, $Q^\varepsilon$ satisfies the same constraints.
	Moreover,
	$
	Q^\varepsilon(\omega)>0
	$ for all $\omega\in\Omega$
	and hence $Q^\varepsilon\in\mathcal M'(Q)$.
	Furthermore,
	\[
	E_{Q^\varepsilon}[F^*]
	=(1-\varepsilon)E_Q[F^*]
	+\varepsilon E_{Q^0}[F^*]
	\longrightarrow E_Q[F^*]
	\]
	as $\varepsilon\downarrow0$.
	Consequently,
	\[
	\max_{Q\in\overline{\mathcal M}'(Q)}
	E_Q[F^*]
	=
	\sup_{Q\in\mathcal M'(Q)}E_Q[F^*].
	\]
	Combining the above equalities gives
	\[
	\Pi^+(F)
	=
	\sup_{Q\in\mathcal M'(Q)}E_Q[F^*],
	\]
	which completes the proof.
\end{proof}

\begin{remark}
	Theorem \ref{theo:suphedge} gives the dual representation of the superhedging price in the finite-state market.
	The supremum over strictly positive martingale measures in $\mathcal M'(Q)$ equals the maximum over the full family $\overline{\mathcal M}'(Q)$, which also allows zero probabilities at terminal states.
	This is the finite-state superhedging duality; a related result in nondominated discrete-time models is given in \citet{Bouchard15}.
\end{remark}

\subsection{Superhedging with short sales prohibitions under model uncertainty}

We now consider the superhedging problem under short sales prohibitions, where the admissible risky-asset positions satisfy $\hat h_m(t)\geq0$ for $m=1,\ldots,M$ and $t=1,\ldots,T$.
By Theorem \ref{theo:fundamental short multi}, the no-arbitrage condition is equivalent to the existence of a strictly positive supermartingale measure.
We collect all such measures in $\tilde{\mathcal M}'(Q)$.
The nonnegativity constraints on risky-asset positions lead to supermartingale inequalities in the dual problem.
The prescribed family $\mathcal Q$ introduced above provides a robust valuation under the strong risk neutral nonlinear expectation condition, whereas the exact constrained superhedging price is determined by the full dual family of supermartingale measures.
We establish this dual representation below.

\begin{theorem}[\textbf{Superhedging theorem under model uncertainty and short sales prohibitions}]
\label{theo:suphedge short}
In multi-period, we assume that there is no-arbitrage under model uncertainty and the short-sales prohibition. 
If $F^*$ is super-replicable, then its constrained superhedging price is given by
\begin{align}
\label{eq:superhedge}
\hat\Pi(F):=&\min\{\hat x\in\mathbb R:\exists\ \hat H\ge0\ \text{such that}\ \hat x+\hat H\cdot S^*(T)\ge F^*,\ \mathcal P\text{-q.s.}\}\\
=&\sup_{Q\in\tilde{\mathcal M}'(Q)}E_Q[F^*],\notag
\end{align}
where $\tilde{\mathcal M}'(Q)$ denotes the set of risk neutral probability measures defined in \eqref{eq:M~'Q}.
\end{theorem}
\begin{proof}
For every $Q\in\tilde{\mathcal M}'(Q)$ and every predictable nonnegative strategy $\hat H$,
\[
E_Q[\hat H\cdot S^*(T)]
=\sum_{u,m}E_Q\!\left[\hat h_m(u)E_Q[\Delta S_m^*(u)\mid\mathcal F_{u-1}]\right]\le0.
\]
Thus each constrained superhedge satisfies $\hat x\ge E_Q[F^*]$, and
\[
\hat\Pi(F)\ge\sup_{Q\in\tilde{\mathcal M}'(Q)}E_Q[F^*].
\]

The constrained primal problem is a finite-dimensional linear program.
Let $\mathcal N_t$ denote the set of nodes at time $t$, and let $\Omega(n)$ denote the set of terminal states whose paths pass through node $n$.
For $\omega\in\Omega$, write $n_t(\omega)$ for the node visited by the path $\omega$ at time $t$.
By the support assumption on $\mathcal P$, the superhedging inequality holds statewise.
Thus, the primal problem can be written as
\[
\begin{aligned}
	\min_{\hat x,\{\hat h_m(t,n)\}}\quad
	&\hat x\\
	\text{subject to}\quad
	&\hat x+
	\sum_{t=1}^{T}\sum_{m=1}^{M}
	\hat h_m(t,n_{t-1}(\omega))
	\Delta S_m^*(t,\omega)
	\geq F^*(\omega),
	\quad \forall\omega\in\Omega,\\
	&\hat h_m(t,n)\geq0,
	\quad n\in\mathcal N_{t-1},\
	t=1,\ldots,T,\quad m=1,\ldots,M.
\end{aligned}
\]
The dual variables are nonnegative weights $q_\omega$ assigned to the terminal states.
The dual constraint corresponding to the initial capital $\hat x$ is
$
\sum_{\omega\in\Omega}q_\omega=1.
$
Since the risky-asset positions are nonnegative, the dual constraints corresponding to $\hat h_m(t,n)$ are
\[
\sum_{\omega\in\Omega(n)}
q_\omega\Delta S_m^*(t+1,\omega)\leq0,
\quad
n\in\mathcal N_t,\quad
t=0,\ldots,T-1,\quad
m=1,\ldots,M.
\]
For a probability measure $Q$ on $\Omega$, these inequalities are the supermartingale flow constraints on the event tree.
Denote by $\overline{\tilde{\mathcal M}}'(Q)$ the family of probability measures satisfying
\[
\begin{aligned}
	\overline{\tilde{\mathcal M}}'(Q)
	=\Biggl\{Q:\;&
	Q(\omega)\geq0,\quad \forall\omega\in\Omega,
	\quad
	\sum_{\omega\in\Omega}Q(\omega)=1,\\
	&\sum_{\omega\in\Omega(n)}
	Q(\omega)\Delta S_m^*(t+1,\omega)\leq0,\quad
	n\in\mathcal N_t,\quad
	t=0,\ldots,T-1,\quad
	m=1,\ldots,M
	\Biggr\}.
\end{aligned}
\]
For any $Q\in\overline{\tilde{\mathcal M}}'(Q)$ and any node $n$ with $Q(n)>0$, where
\[
Q(n):=\sum_{\omega\in\Omega(n)}Q(\omega),
\]
the corresponding flow inequality is equivalent to
$
E_Q[\Delta S_m^*(t+1)\mid\mathcal F_t](n)\leq0.
$
If $Q(n)=0$, then $Q(\omega)=0$ for every $\omega\in\Omega(n)$, and the corresponding flow inequality is automatically satisfied.
Thus, $\overline{\tilde{\mathcal M}}'(Q)$ is the closed family of probability measures satisfying the supermartingale flow constraints, where zero probabilities at terminal states are allowed.
The dual objective is
\[
\sum_{\omega\in\Omega}q_\omega F^*(\omega)
=E_Q[F^*].
\]
The primal problem is feasible by taking $\hat H=0$ and
$
\hat x=\max_{\omega\in\Omega}F^*(\omega).
$
Moreover, by Theorem \ref{theo:fundamental short multi}, there exists a strictly positive supermartingale measure
$Q^0\in\tilde{\mathcal M}'(Q)$.
For every feasible superhedging strategy, the supermartingale property of $Q^0$ gives
\[
\hat x\geq E_{Q^0}[F^*].
\]
Hence, the primal problem is bounded below. The dual feasible set is nonempty, since it contains $Q^0$, and is a closed subset of the finite-dimensional probability simplex. It is therefore compact.
Strong linear-programming duality yields
\[
\hat\Pi(F)
=
\max_{Q\in\overline{\tilde{\mathcal M}}'(Q)}
E_Q[F^*].
\]
We now show that the maximum over $\overline{\tilde{\mathcal M}}'(Q)$ equals the supremum over strictly positive supermartingale measures.
For any $Q\in\overline{\tilde{\mathcal M}}'(Q)$ and any
$\varepsilon\in(0,1)$, define
\[
Q^\varepsilon
:=
(1-\varepsilon)Q+\varepsilon Q^0.
\]
Since the supermartingale flow inequalities are linear in the terminal-state probabilities, $Q^\varepsilon$ satisfies the same inequalities.
Moreover,
$
Q^\varepsilon(\omega)>0,
$ for all $\omega\in\Omega$.
and hence $Q^\varepsilon$ is a strictly positive one-period supermartingale measure.
By the tower property, it satisfies the supermartingale condition for all $0\leq t\leq u\leq T$, so that
$Q^\varepsilon\in\tilde{\mathcal M}'(Q)$.
Furthermore,
\[
E_{Q^\varepsilon}[F^*]
=
(1-\varepsilon)E_Q[F^*]
+
\varepsilon E_{Q^0}[F^*]
\longrightarrow
E_Q[F^*]
\]
as $\varepsilon\downarrow0$.
Consequently,
\[
\max_{Q\in\overline{\tilde{\mathcal M}}'(Q)}
E_Q[F^*]
=
\sup_{Q\in\tilde{\mathcal M}'(Q)}
E_Q[F^*].
\]
Combining this equality with the dual representation above gives
\[
\hat\Pi(F)
=
\sup_{Q\in\tilde{\mathcal M}'(Q)}
E_Q[F^*],
\]
which proves equation \eqref{eq:superhedge}.
\end{proof}

\begin{remark}
	Theorems \ref{theo:suphedge} and \ref{theo:suphedge short} involve different dual families.
	In the unrestricted market, the dual constraints are martingale equalities, whereas under short sales prohibitions, they are supermartingale inequalities.
	
	Now suppose that a prescribed family $\mathcal Q$ satisfies the strong risk neutral nonlinear expectation condition in Definition \ref{de:risk neutral strong expectation multi}.
	Then every $Q\in\mathcal Q$ is a supermartingale measure, and hence
	$\mathcal Q\subseteq\overline{\tilde{\mathcal M}}'(Q)$.
	Consequently,
	\[
	\begin{aligned}
		\sup_{Q\in\mathcal Q}E_Q[F^*]
		&\leq
		\max_{Q\in\overline{\tilde{\mathcal M}}'(Q)}
		E_Q[F^*]\\
		&=\hat\Pi(F)
		=\sup_{Q\in\tilde{\mathcal M}'(Q)}E_Q[F^*].
	\end{aligned}
	\]
	Thus, the upper expectation over the prescribed family $\mathcal Q$ provides a lower bound for the exact constrained superhedging price.
	Equality holds when the selected family attains the same supremum as the complete dual family for the claim under consideration.
\end{remark}

\subsection{Backward linear-programming implementation}

In the finite-state event-tree setting, the superhedging problems with and without short-sales prohibitions can be solved within a unified framework by nodewise linear programming and backward induction. 
For each node $n$ at time $t$, let $Y_t(n)$ denote the minimum discounted wealth required at that node to superhedge the remaining claim. At each terminal node $\omega\in\Omega$, set
\[
Y_T(\omega)=F^*(\omega).
\]
The same notation $Y_t$ is used for the nodewise superhedging value in the unrestricted and short-sales constrained markets, with the corresponding trading constraint imposed in each case. The backward procedure is carried out as follows.

{
	\renewcommand{\labelenumi}{\textbf{Step \theenumi:}}
	\begin{enumerate}
		
		\item
		Fix a nonterminal node $n$ at time $t$ and suppose that the continuation values
		$Y_{t+1}(n')$ are known for all
		$n'\in\operatorname{succ}(n)$. Let $v_t(n)$ denote a candidate discounted wealth at node $n$, and let
		\[
		h(t+1,n)
		=
		\bigl(
		h_1(t+1,n),\ldots,h_M(t+1,n)
		\bigr)^\top
		\]
		denote the risky-asset position selected at node $n$ and held over the period
		$(t,t+1)$. 
		The nodewise superhedging problem is
		\[
		Y_t(n)
		:=
		\min_{v_t(n),\,h(t+1,n)}
		v_t(n)
		\]
		subject to
		\[
		v_t(n)
		+
		\sum_{m=1}^{M}
		h_m(t+1,n)
		\Bigl(
		S_m^*(t+1,n')-S_m^*(t,n)
		\Bigr)
		\geq
		Y_{t+1}(n'),
		\qquad
		n'\in\operatorname{succ}(n).
		\]
		In the unrestricted market,
		$
		h_m(t+1,n)\in\mathbb R,
		\ 
		m=1,\ldots,M.
		$
		Under short sales prohibitions, impose instead
		$
		h_m(t+1,n)\geq0,
		\ 
		m=1,\ldots,M.
		$
		
		\item
		The dual variables associated with the successor-state constraints can be
		normalized as a local probability kernel
		$q_t(\cdot\mid n)$ satisfying
		\[
		q_t(n'\mid n)\geq0,
		\qquad
		n'\in\operatorname{succ}(n),
		\]
		and
		\[
		\sum_{n'\in\operatorname{succ}(n)}
		q_t(n'\mid n)
		=
		1.
		\]
		The dual objective is
		\[
		\max_{q_t(\cdot\mid n)}
		\sum_{n'\in\operatorname{succ}(n)}
		q_t(n'\mid n)Y_{t+1}(n').
		\]
		
		In the unrestricted market, the dual kernel satisfies the nodewise martingale
		conditions
		\[
		\sum_{n'\in\operatorname{succ}(n)}
		q_t(n'\mid n)
		S_m^*(t+1,n')
		=
		S_m^*(t,n),
		\qquad
		m=1,\ldots,M.
		\]
		Under short sales prohibitions, these equalities are replaced by the
		supermartingale inequalities
		\[
		\sum_{n'\in\operatorname{succ}(n)}
		q_t(n'\mid n)
		S_m^*(t+1,n')
		\leq
		S_m^*(t,n),
		\qquad
		m=1,\ldots,M.
		\]
		Thus, the dual feasible kernels satisfy the local martingale conditions in the unrestricted market and the local supermartingale conditions under short sales prohibitions.
		
		\item
		Starting from
		\[
		Y_T(\omega)=F^*(\omega),
		\qquad
		\omega\in\Omega,
		\]
		repeat the preceding nodewise optimization backward for
		$
		t=T-1,T-2,\ldots,0.
		$
		At each node, the primal problem yields the minimum discounted wealth required to superhedge the remaining claim and an optimal risky-asset position, while the dual problem yields an optimizing conditional probability kernel.
		At the initial node $n_0$, we obtain
		\[
		Y_0(n_0)=\Pi^+(F)
		\]
		in the unrestricted market, and
		\[
		Y_0(n_0)=\hat\Pi(F)
		\]
		under short sales prohibitions.
		The local conditional kernels can be multiplied along each path to obtain the corresponding global dual probability measure.
	\end{enumerate}
}

For numerical optimization, the local dual kernels are allowed to have zero components, so that the dual feasible sets are closed.
The resulting global dual measures may therefore assign zero probabilities to some terminal states.
Under the no-arbitrage conditions of Theorems \ref{theo:fundamental multi} and \ref{theo:fundamental short multi}, there exist strictly positive martingale and supermartingale measures, respectively.
By mixing any boundary dual measure with such a strictly positive measure, we obtain strictly positive admissible measures whose expectations converge to that of the original measure.
Consequently, allowing zero components in the local probability kernels does not change the optimal values given by the dual representations in Theorems \ref{theo:suphedge} and \ref{theo:suphedge short}.

For the empirical implementation in Section \ref{sec:empirical}, the discounted conditions above can equivalently be written in terms of one-period gross returns.
In the one-risky-asset setting used there, let $g_t$ denote the vector of gross risky returns across the successor states, and let $R_{t+1}=S_0(t+1)/S_0(t)$ denote the gross risk-free return over the period $(t,t+1)$.
The local martingale and supermartingale conditions then become
$
q_t^\top g_t=R_{t+1}
$
and
$
q_t^\top g_t\leq R_{t+1},
$
respectively.
Thus, the probability constraints used in Section \ref{sec:empirical} are the gross-return counterparts of the discounted-price conditions above.

For an empirically constructed ambiguity family, probability bounds can also be incorporated into the local optimization problems when the family is specified through nodewise conditional probability constraints.
In this case, the corresponding local kernels can be selected independently across nodes, and the resulting family is rectangular.
The lower and upper nonlinear expectations are then obtained by taking the infimum and supremum of the conditional expectations over the admissible kernels, respectively.
Rolling windows are used to estimate the probability bounds, while the nonlinear expectations are defined by the resulting family of probability measures.
This procedure is used in the empirical illustration of Section \ref{sec:empirical}.

\section{Numerical illustration of pricing and superhedging using real data}
\label{sec:empirical}

\subsection{Data, calibration, and state construction}

To connect the preceding theory with observed asset dynamics, we use monthly S\&P 500 price-index data\footnote{S\&P 500 historical price-index data are obtained from Yahoo Finance and the Federal Reserve Bank of St. Louis (FRED), \url{https://fred.stlouisfed.org/series/SP500}.}, with one completed month-end observation retained for each calendar month before returns are calculated. 
We write $S_t$ for the month-end index level and $G_t=S_t/S_{t-1}$ for the monthly gross return.
Monthly observations are used so that each trading period corresponds to one month, consistently with the two-period option setting adopted below. The resulting series covers January 1990 through August 2026, with the starting date providing sufficient historical observations for the initial calibration. For each valuation date $t$, only historical observations available before $t$ are used; no future return enters the calibration.

The claims are synthetic two-month at-the-money European put and call options with $K=S_t$,
\[
F_{t+2}^{\rm put}=(K-S_{t+2})^+,
\qquad
F_{t+2}^{\rm call}=(S_{t+2}-K)^+.
\]
The next two realized monthly index levels are reserved exclusively for the subsequent hedge evaluation. After the 120-month initialization and the two-month evaluation requirement, the valuation sample contains 317 dates from February 2000 to June 2026. The baseline annual risk-free rate is fixed at $2\%$ and converted to the monthly gross return
$
R=(1.02)^{1/12}.
$
This constant-rate specification keeps the numerical comparison focused on the effects of model uncertainty and trading constraints and is used as a benchmark assumption rather than as a maturity-matched market rate.
The underlying is a price index, so dividends are not included. Transaction costs are set to zero.
At each valuation date, the 5th, 50th, and 95th empirical quantiles of the preceding 120 monthly gross returns define the lower, median, and upper return states,
\[
g_t=(g_{d,t},g_{m,t},g_{u,t}),
\qquad
g_{d,t}<g_{m,t}<g_{u,t},
\qquad
g_{d,t}<R<g_{u,t}.
\]
The empirical quantiles are used directly without ex post adjustment, and these conditions hold throughout the sample. The two-period terminal grid is
\[
S_{t+2}^{ij}=S_tg_{i,t}g_{j,t},
\qquad i,j\in\{d,m,u\}.
\]

For each of the 36-, 60-, and 120-month frequency windows,
we assign historical returns to the three states using
the midpoints between adjacent state multipliers. Set
\[
c_{1,t}=\frac{g_{d,t}+g_{m,t}}{2},
\qquad
c_{2,t}=\frac{g_{m,t}+g_{u,t}}{2}.
\]
The three intervals are
\[
I_{d,t}=(-\infty,c_{1,t}),\qquad
I_{m,t}=[c_{1,t},c_{2,t}),\qquad
I_{u,t}=[c_{2,t},\infty).
\]
Let $N_{k,t}^{(L)}$ be the number of returns in $I_{k,t}$
among the preceding $L$ monthly observations, where
$L\in\{36,60,120\}$. We use the Jeffreys correction
\[
p_{k,t}^{(L)}
=
\frac{N_{k,t}^{(L)}+1/2}{L+3/2},
\qquad k\in\{d,m,u\}.
\]
Each resulting frequency vector is exponentially tilted
to satisfy the martingale condition:
\[
q_{k,t}^{(L)}
=
\frac{
	p_{k,t}^{(L)}
	\exp\{\lambda_t^{(L)}(g_{k,t}-R)\}
}{
	\sum_{j\in\{d,m,u\}}
	p_{j,t}^{(L)}
	\exp\{\lambda_t^{(L)}(g_{j,t}-R)\}
},
\]
where $\lambda_t^{(L)}$ is chosen so that
\[
\sum_{k\in\{d,m,u\}}q_{k,t}^{(L)}g_{k,t}=R.
\]
The solution is unique since $g_{d,t}<R<g_{u,t}$
and all three frequencies are positive.
Let $q_t^{36}$, $q_t^{60}$, and $q_t^{120}$ denote the resulting kernels. The 60-month kernel is used as the single-model benchmark.
To represent model uncertainty, we construct a data-supported family of probability kernels from these three estimates. For a probability-bound expansion parameter $\delta\geq0$, define
\[
\mathcal A_t(\delta)
=
\left\{
q\geq0:
\mathbf 1^\top q=1,\ 
\underline q_t(\delta)\leq q\leq\bar q_t(\delta)
\right\},
\]
where the lower and upper bounds are obtained from the componentwise minima and maxima of the three kernels and are further expanded by $\delta$. A larger $\delta$ produces a wider family of admissible probability models. The baseline analysis uses $\delta=0.04$.

To illustrate the effect of short-sale constraints, we compare two trading regimes on the same calibrated market. In the unrestricted regime, the risky-asset position may take arbitrary real values, whereas in the constrained regime we impose $h_t\geq0$. This restriction is introduced for the purpose of illustrating the theoretical effect of short-sale constraints and is not intended to represent the actual short-sale regulation of the S\&P 500 market. The corresponding data-supported dual families are
\[
\mathcal N_t(\mathcal A)
=
\{q\in\mathcal A_t:q^\top g_t=R\},
\qquad
\mathcal S_t(\mathcal A)
=
\{q\in\mathcal A_t:q^\top g_t\leq R\}.
\]
The same one-period probability family is available independently at each node, so the resulting two-period family is rectangular by construction, consistently with Section \ref{sec:multi}.

For comparison, the complete local dual sets are obtained by removing the empirical probability bounds. To relate the calibrated family $\mathcal A_t$ to the weak and strong risk-neutral nonlinear expectation conditions, define
\[
d_t^-=
\min_{q\in\mathcal A_t}(q^\top g_t-R),
\qquad
d_t^+=
\max_{q\in\mathcal A_t}(q^\top g_t-R).
\]
The weak condition holds when $d_t^-\leq0$, whereas the strong condition holds when $d_t^+\leq0$. Since $\mathcal A_t$ contains the three martingale kernels, the weak condition holds by construction. In the sample, $d_t^+>0$ at every valuation date, so $\mathcal A_t$ itself does not satisfy the strong condition. This does not imply arbitrage, since the strong condition is sufficient but not necessary for no-arbitrage. Under the short-sale constraint, the data-supported valuation is computed over the supermartingale subfamily $\mathcal S_t(\mathcal A)$, while the exact superhedging price is obtained from the complete supermartingale dual family.

\subsection{Pricing results and primal-dual verification}

We examine the effects of model uncertainty and short-sale constraints on the valuations and superhedging costs of the synthetic two-month at-the-money put and call options.
Table~\ref{tab:empiricalpricing} reports summary statistics for the 317 valuation dates, and Figure~\ref{fig:mean-prices} displays selected mean prices.
The benchmark uses the 60-month martingale kernel.
The data-supported martingale bounds and supermartingale upper value are computed by backward optimization over the rectangular families generated by $\mathcal N_t(\mathcal A)$ and $\mathcal S_t(\mathcal A)$, respectively.
The unrestricted and no-short superhedging prices use the complete martingale and supermartingale dual families, respectively, and equal the corresponding minimum primal costs by Theorems~\ref{theo:suphedge} and~\ref{theo:suphedge short}.
Prices are expressed as percentages of the contemporaneous index level.
Mean and Median summarize price levels across dates, Std. measures their dispersion in percentage points, and the 5\% and 95\% columns report empirical percentiles.
These statistics describe variation across rolling calibrations, not valuation uncertainty across models at a fixed date; the percentiles are not confidence limits.

\begin{table}[!ht]
	\centering
	\caption{Rolling two-period put and call valuations from February 2000 to June 2026}
	\label{tab:empiricalpricing}
	\begingroup
	\footnotesize
	\linespread{1.25}\selectfont
	\renewcommand{\arraystretch}{1.00}
	\setlength{\tabcolsep}{3pt}
	\begin{tabular*}{\textwidth}{@{\extracolsep{\fill}}llrrrrr@{}}
		\toprule
		Claim & Quantity & Mean & Median & Std. & 5\% & 95\% \\
		\midrule
		Put & Single-model benchmark
		& 2.405 & 2.466 & 0.423 & 1.678 & 2.938 \\
		Put & Data-supported martingale lower
		& 2.030 & 2.124 & 0.394 & 1.211 & 2.547 \\
		Put & Data-supported martingale upper
		& 2.759 & 2.744 & 0.284 & 2.278 & 3.233 \\
		Put & \begin{tabular}[c]{@{}l@{}}
			Unrestricted superhedging price\\
		\end{tabular}
		& 3.583 & 3.606 & 0.325 & 3.062 & 4.248 \\
		Put & Data-supported supermartingale upper
		& 3.470 & 3.471 & 0.387 & 2.858 & 4.157 \\
		Put & \begin{tabular}[c]{@{}l@{}}
			No-short superhedging price\\
		\end{tabular}
		& 13.648 & 13.981 & 1.674 & 9.913 & 16.346 \\
		\midrule
		Call & Single-model benchmark
		& 2.734 & 2.796 & 0.423 & 2.007 & 3.268 \\
		Call & Data-supported martingale lower
		& 2.359 & 2.454 & 0.394 & 1.540 & 2.877 \\
		Call & Data-supported martingale upper
		& 3.088 & 3.074 & 0.284 & 2.607 & 3.562 \\
		Call & \begin{tabular}[c]{@{}l@{}}
			Unrestricted superhedging price\\
		\end{tabular}
		& 3.913 & 3.935 & 0.325 & 3.392 & 4.578 \\
		Call & Data-supported supermartingale upper
		& 3.088 & 3.074 & 0.284 & 2.607 & 3.562 \\
		Call & \begin{tabular}[c]{@{}l@{}}
			No-short superhedging price\\
		\end{tabular}
		& 3.913 & 3.935 & 0.325 & 3.392 & 4.578 \\
		\bottomrule
	\end{tabular*}
	\par\smallskip
	\begin{minipage}{\textwidth}
		\footnotesize
		Notes: Prices are percentages of the contemporaneous index level;
		Std. is measured in percentage points.
	\end{minipage}
	\endgroup
\end{table}

Allowing for model uncertainty yields a date-specific valuation interval containing the benchmark, since $q_t^{60}\in\mathcal N_t(\mathcal A)$.
For the put, the mean benchmark is $2.405\%$, while the mean martingale lower and upper values are $2.030\%$ and $2.759\%$; the corresponding call values are $2.734\%$, $2.359\%$, and $3.088\%$.
These averages summarize date-specific intervals rather than a single interval applying throughout the sample.
The mean data-supported upper values are nevertheless below the corresponding mean superhedging prices.
For example, the mean put martingale upper value is $2.759\%$, compared with an unrestricted superhedging cost of $3.583\%$; the supermartingale upper value is $3.470\%$, compared with a no-short cost of $13.648\%$.
Superhedging requires terminal wealth to cover the payoff on every path of the calibrated tree.
The empirical probability bounds restrict how these paths are weighted but do not remove any of them, so they do not relax the pathwise hedging constraints.
The complete dual optimizes over additional martingale or supermartingale measures excluded by these bounds and can therefore yield a higher value.
Consequently, a data-supported upper value need not be sufficient to finance a superhedge; equality requires the selected family to yield the same supremum as the complete dual family.

Short-sale constraints have a pronounced effect on the put but do not bind for the call.
In Table~\ref{tab:empiricalpricing} and Figure~\ref{fig:mean-prices}, the mean put superhedging price increases from $3.583\%$ to $13.648\%$, whereas the mean call price remains $3.913\%$.
A short stock position helps finance the increasing put payoff when the index falls.
Without short selling, the two-down-move path determines the required initial capital: it has the largest put payoff, while nonnegative stock holdings generate nonpositive discounted gains on this path because $g_{d,t}<R$.
Initial capital must therefore be at least the discounted maximum payoff.
Investing that amount in the risk-free asset covers the payoff on every tree path, so a cash-only strategy is optimal in each window.
For the call, a long stock position gains as the payoff increases, and the computed unrestricted optimal holdings are nonnegative at every trading node.
The same strategy is therefore feasible under the constraint, giving identical superhedging prices.
The put cost increase is also reflected in the median ($13.981\%$ versus $3.606\%$), so the high constrained mean is not solely attributable to a few large observations.
The no-short put prices have a larger standard deviation ($1.674$ versus $0.325$ percentage points) and a wider 5th-95th percentile range ($[9.913\%,16.346\%]$ versus $[3.062\%,4.248\%]$).
With $K=S_t$ and fixed $R$, the variation in the normalized cash-only price is determined by changes in the calibrated down-state return $g_{d,t}$.

\begin{figure}[!ht]
	\centering
	\begin{tikzpicture}
		\begin{axis}[
			width=0.96\textwidth,
			height=7.6cm,
			ybar=1.5pt,
			bar width=29pt,
			xmin=-0.60, xmax=1.60,
			ymin=0, ymax=15,
			enlarge x limits=false,
			xtick={0,1},
			xticklabels={Put,Call},
			ytick={0,3,6,9,12,15},
			ylabel={Mean price (\% of spot)},
			tick label style={font=\small},
			label style={font=\small},
			tick align=outside,
			ymajorgrids=true,
			grid style={gray!18, line width=0.25pt},
			axis line style={black, line width=0.45pt},
			area legend,
			legend style={
				at={(0.5,1.04)}, anchor=south,
				legend columns=2,
				font=\footnotesize,
				draw=none, fill=none,
				column sep=8pt, row sep=2pt
			},
			legend cell align=left,
			nodes near coords,
			every node near coord/.append style={
				font=\scriptsize, text=black,
				/pgf/number format/.cd,
				fixed, precision=3, zerofill
			}
			]
			\addplot[draw=blue, fill=blue!30!white]
			coordinates {(0,2.405) (1,2.734)};
			\addplot[draw=red, fill=red!30!white]
			coordinates {(0,2.759) (1,3.088)};
			\addplot[draw=brown!60!black, fill=brown!30!white]
			coordinates {(0,3.583) (1,3.913)};
			\addplot[draw=black, fill=gray]
			coordinates {(0,13.648) (1,3.913)};
			\legend{
				Single-model benchmark,
				Data-supported martingale upper,
				Unrestricted superhedging price,
				No-short superhedging price
			}
		\end{axis}
	\end{tikzpicture}
	\caption{Mean option valuations and superhedging prices}
	\label{fig:mean-prices}
\end{figure}

Finally, all 317 valuation dates are solved successfully.
The maximum absolute gaps between the separately solved primal and complete dual problems are approximately $1.88\times10^{-12}$ without short-sale constraints and $9.66\times10^{-13}$ with the constraint.
These gaps provide a numerical consistency check on the implementation of the finite-tree dualities in Section \ref{sec:hedging}.
Performance on realized paths is examined in the next subsection.

\subsection{Out-of-sample hedge evaluation}

We evaluate the superhedging strategies from the preceding subsection along the subsequent observed index paths. Initial capital and positions are obtained from the calibrated two-period tree. At $t+1$, the adaptive extension retains $g_t$, rescales the remaining one-period tree by the observed $S_{t+1}$, and solves the local superhedging LP to update the stock position. The risk-free holding is adjusted to preserve self-financing; $S_{t+2}$ is used only for terminal evaluation.
Table~\ref{tab:hedgingperformance} summarizes the 317 rolling windows. Mean initial cost is the average capital committed at $t$. With terminal portfolio wealth $V(t+2)$, hedging deficit and surplus are $(F_{t+2}-V(t+2))^+$ and $(V(t+2)-F_{t+2})^+$, respectively. Their means include zero observations, and all monetary quantities are normalized as percentages of $S_t$ before averaging.
Deficit frequency is the percentage of windows with a positive deficit.
In range is the percentage of windows in which both $S_{t+1}/S_t$ and $S_{t+2}/S_{t+1}$ lie in $[g_{d,t},g_{u,t}]$, without requiring equality to a discrete tree state.
The final column reports the deficit frequency conditional on these windows.

\begin{table}[!ht]
	\centering
	\caption{Out-of-sample performance of the extended superhedging strategies}
	\label{tab:hedgingperformance}
	\begingroup
	\small
	\linespread{1.05}\selectfont
	\renewcommand{\arraystretch}{1.12}
	\setlength{\tabcolsep}{2.5pt}
	\begin{tabular*}{\textwidth}{@{\extracolsep{\fill}}cccccccc@{}}
		\toprule
		Claim & Strategy
		& \begin{tabular}[c]{@{}c@{}}
			Mean initial\\cost
		\end{tabular}
		& \begin{tabular}[c]{@{}c@{}}
			Mean hedging\\deficit
		\end{tabular}
		& \begin{tabular}[c]{@{}c@{}}
			Deficit\\frequency
		\end{tabular}
		& \begin{tabular}[c]{@{}c@{}}
			Mean\\surplus
		\end{tabular}
		& In range
		& \begin{tabular}[c]{@{}c@{}}
			Deficit frequency\\in range
		\end{tabular} \\
		\midrule
		Put & Unrestricted
		& 3.583 & 0.207 & 18.93 & 1.350 & 78.55 & 0.00 \\
		Put & No short selling
		& 13.648 & 0.135 & 2.52 & 11.877 & 78.55 & 0.00 \\
		\midrule
		Call & Unrestricted
		& 3.913 & 0.207 & 18.93 & 1.350 & 78.55 & 0.00 \\
		Call & No short selling
		& 3.913 & 0.207 & 18.93 & 1.350 & 78.55 & 0.00 \\
		\bottomrule
	\end{tabular*}
	\par\smallskip
	\begin{minipage}{\textwidth}
		\footnotesize
		Notes: Mean initial cost, mean hedging deficit, and mean surplus are
		averages over all 317 windows of amounts expressed as percentages of
		$S_t$; zero deficits and surpluses are included.
		Deficit frequencies and In range are percentages of windows.
		In range requires both realized monthly gross returns to lie in
		$[g_{d,t},g_{u,t}]$ and does not require them to equal tree states.
		The final column uses only the in-range windows as its denominator.
	\end{minipage}
	\endgroup
\end{table}

Over the full sample, the unrestricted put and call strategies each have a deficit frequency of $18.93\%$ and a mean deficit of $0.207\%$ of spot.
Thus, exact superhedging on the calibrated tree does not provide complete protection along the observed paths.
The adaptive extension adjusts the position to the observed intermediate index level but retains the calibrated return states.
In $78.55\%$ of the windows, both realized returns remain within the calibrated range, and no deficit is observed for any strategy.
This pattern is consistent with the convexity of the payoffs: one-period portfolio wealth is affine in the terminal index level, so covering the payoffs at both endpoints also covers intermediate levels.
Beyond the endpoints, the local LP imposes no such coverage requirement.
All recorded deficits occur in windows with at least one return outside the calibrated range, indicating exposure to movements beyond the estimated bounds.
The zero conditional frequency is an empirical result for the evaluated windows and the specified extension rule, rather than a general guarantee for arbitrary paths outside the discrete tree.

Short-sale constraints change this performance substantially for the put.
The mean initial cost increases from $3.583\%$ to $13.648\%$ of spot, while the deficit frequency falls from $18.93\%$ to $2.52\%$ and the mean deficit falls from $0.207\%$ to $0.135\%$.
As established in the preceding subsection, the no-short put hedge holds only the risk-free asset.
Its initial reserve covers the largest put payoff on the calibrated tree; on a realized path, a deficit occurs only if the payoff exceeds the accumulated cash.
The mean surplus is $11.877\%$, compared with $1.350\%$ for the unrestricted hedge.
The lower deficit frequency therefore reflects greater initial funding, with much of the cash remaining after settlement.
For the call, the unrestricted stock positions are nonnegative both initially and under the one-period rebalancing rule, so the short-sale constraint does not exclude the strategy.
Initial costs and all reported performance measures consequently coincide under the two regimes.
The effect on realized hedge performance is therefore payoff-dependent.

\subsection{Sensitivity to probability ambiguity}

We assess the sensitivity of the data-supported valuations to the
probability-bound expansion parameter $\delta$, using $0.00$, $0.02$,
$0.04$, and $0.06$ around the baseline value of $0.04$. At each valuation
date, the return states, the three calibrated martingale kernels, and
the risk-free rate are held fixed. Table~\ref{tab:sensitivity} reports
the results, and Figure~\ref{fig:sensitivity} displays the three
upper-value series. 
For each value of $\delta$, the valuation width over $\mathcal A_t(\delta)$ is first computed as the difference between the two-period upper and lower valuations over the full data-supported probability family, before imposing either drift condition.
The martingale lower and upper values are then computed over $\mathcal N_t(\mathcal A)$, while the supermartingale upper value is computed over $\mathcal S_t(\mathcal A)$. 
All valuation statistics are averaged across the 317 dates after normalization by the contemporaneous index level. 

\begin{table}[!ht]
	\centering
	\caption{Valuation sensitivity to the probability-bound expansion}
	\label{tab:sensitivity}
	\begingroup
	\small
	\linespread{1.05}\selectfont
	\renewcommand{\arraystretch}{1.12}
	\setlength{\tabcolsep}{3pt}
	\begin{tabular*}{\textwidth}{@{\extracolsep{\fill}}cccccc@{}}
		\toprule
		Claim & $\delta$
		& \begin{tabular}[c]{@{}c@{}}Valuation width\\over $\mathcal A_t(\delta)$\end{tabular}
		& \begin{tabular}[c]{@{}c@{}}Martingale\\lower\end{tabular}
		& \begin{tabular}[c]{@{}c@{}}Martingale\\upper\end{tabular}
		& \begin{tabular}[c]{@{}c@{}}Supermartingale\\upper\end{tabular} \\
		\midrule
		Put  & 0.00 & 0.869 & 2.173 & 2.645  & 2.875  \\
		Put  & 0.02 & 1.415 & 2.101 & 2.702  & 3.169  \\
		Put  & 0.04 & 1.957 & 2.030 & 2.759  & 3.470  \\
		Put  & 0.06 & 2.496 & 1.959 & 2.813  & 3.778  \\
		\midrule
		Call & 0.00 & 1.096 & 2.503 & 2.974  & 2.974  \\
		Call & 0.02 & 1.682 & 2.431 & 3.032  & 3.032  \\
		Call & 0.04 & 2.259 & 2.359 & 3.088  & 3.088  \\
		Call & 0.06 & 2.829 & 2.289 & 3.142  & 3.142  \\
		\bottomrule
	\end{tabular*}
	\par\smallskip
	\begin{minipage}{\textwidth}
		\footnotesize
		Notes: All entries are means across 317 valuation dates.
		Lower and upper valuations are expressed as percentages of the contemporaneous index level, while valuation widths are measured in percentage points.
		The valuation width over $\mathcal A_t(\delta)$ is computed before imposing the martingale or supermartingale drift condition and therefore is not the difference between the two martingale columns.
	\end{minipage}
	\endgroup
\end{table}


As $\delta$ increases, the data-supported probability families expand, and the valuation dispersion over $\mathcal A_t(\delta)$ increases accordingly. The mean valuation width rises from $0.869$ to $2.496$ percentage points for the put and from
$1.096$ to $2.829$ percentage points for the call as $\delta$ increases from $0$ to $0.06$.
The martingale valuation intervals widen in the same direction. For the put, the mean martingale lower value decreases from $2.173\%$ to $1.959\%$, while the corresponding upper value increases from $2.645\%$ to $2.813\%$. For the call, the lower value decreases from $2.503\%$ to $2.289\%$, while the upper value increases from $2.974\%$ to $3.142\%$. These results show that a larger probability-bound expansion produces greater valuation dispersion both over the full data-supported model family and within its martingale subfamily.
The nonzero valuation width at $\delta=0$ reflects the dispersion among the three calibrated kernels $q_t^{36}$, $q_t^{60}$, and $q_t^{120}$. Thus, $\delta=0$ means that no additional probability-bound expansion is imposed; it does not correspond to the absence of model uncertainty.

The additional effect of the supermartingale condition differs
between the two payoffs. Over the same range of $\delta$, the mean
put supermartingale upper value rises from $2.875\%$ to $3.778\%$,
an increase of $0.903$ percentage points. In comparison, the put and
call martingale upper values each increase by $0.168$ percentage
points, consistently with put-call parity. Since
$\mathcal N_t(\mathcal A)\subseteq\mathcal S_t(\mathcal A)$,
replacing $q^\top g_t=R$ by $q^\top g_t\leq R$ permits negative
discounted drift and allows greater weight on low-return states,
which benefits the decreasing put payoff. For the increasing call
payoff, this additional freedom does not raise the reported upper
values: the martingale and supermartingale upper values coincide
at all four parameter choices. Figure~\ref{fig:sensitivity}
therefore shows a much larger response for the put supermartingale
upper value than for either martingale upper value.

\begin{figure}[!ht]
	\centering
	\begingroup
	\definecolor{sensblue}{RGB}{63,107,164}
	\definecolor{sensred}{RGB}{190,83,86}
	\definecolor{sensbrown}{RGB}{153,117,70}
	\begin{tikzpicture}
		\begin{axis}[
			width=0.96\textwidth,
			height=7.4cm,
			xmin=-0.003,
			xmax=0.063,
			ymin=2.55,
			ymax=3.95,
			scaled x ticks=false,
			xtick={0,0.02,0.04,0.06},
			ytick={2.6,2.8,3.0,3.2,3.4,3.6,3.8},
			xlabel={Probability-bound expansion parameter $\delta$},
			ylabel={Mean upper value (\% of spot)},
			tick label style={font=\small},
			label style={font=\small},
			xticklabel style={
				/pgf/number format/fixed,
				/pgf/number format/precision=2,
				/pgf/number format/zerofill
			},
			yticklabel style={
				/pgf/number format/fixed,
				/pgf/number format/precision=1,
				/pgf/number format/zerofill
			},
			tick align=outside,
			tick pos=left,
			xmajorgrids=false,
			ymajorgrids=true,
			minor x tick num=0,
			minor y tick num=0,
			grid style={
				gray!15,
				line width=0.25pt
			},
			axis line style={
				black,
				line width=0.45pt
			},
			legend style={
				at={(0.025,0.975)},
				anchor=north west,
				legend columns=1,
				font=\footnotesize,
				draw=none,
				fill=white,
				inner sep=3pt,
				row sep=2pt
			},
			legend cell align=left
			]
			
			\addplot[
			color=sensblue,
			line width=1.1pt,
			mark=*,
			mark size=2.6pt,
			mark options={solid,fill=sensblue}
			]
			coordinates {
				(0.00,2.645)
				(0.02,2.702)
				(0.04,2.759)
				(0.06,2.813)
			};
			
			\addplot[
			color=sensred,
			line width=1.1pt,
			mark=square*,
			mark size=2.6pt,
			mark options={solid,fill=sensred}
			]
			coordinates {
				(0.00,2.875)
				(0.02,3.169)
				(0.04,3.470)
				(0.06,3.778)
			};
			
			\addplot[
			color=sensbrown,
			line width=1.1pt,
			mark=triangle*,
			mark size=3pt,
			mark options={solid,fill=sensbrown}
			]
			coordinates {
				(0.00,2.974)
				(0.02,3.032)
				(0.04,3.088)
				(0.06,3.142)
			};
			
			\legend{
				Put martingale upper,
				Put supermartingale upper,
				Call martingale and supermartingale upper
			}
			
		\end{axis}
	\end{tikzpicture}
	\endgroup
	\caption{Sensitivity of mean data-supported upper values to $\delta$}
	\label{fig:sensitivity}
\end{figure}

The exact superhedging prices are independent of $\delta$, since their complete dual problems are defined on the same calibrated tree without the empirical probability bounds that determine $\mathcal A_t(\delta)$. At every tested value of $\delta$, the mean data-supported upper valuations remain below the corresponding mean superhedging prices reported in Table~\ref{tab:empiricalpricing}. 
Hence, the distinction between data-supported valuation and exact superhedging is not specific to the baseline choice $\delta=0.04$, but persists throughout the sensitivity analysis.


\section{Conclusion}
\label{sec:conclude}

In this paper, we study asset pricing and hedging under model uncertainty in discrete time and finite states. For the single-period model without short-sale restrictions, we characterize no-arbitrage by the existence of a strictly positive risk-neutral probability measure. 
When short sales are prohibited, we obtain the corresponding characterization in terms of strictly positive supermartingale measures and introduce weak and strong risk-neutral nonlinear expectations. 
The existence of a weak risk-neutral nonlinear expectation is necessary for no-arbitrage, whereas the existence of a strong one is sufficient. 
We then extend the analysis to multi-period markets and establish the corresponding fundamental theorems of asset pricing. 
Building on these results, we derive the risk-neutral valuation of replicable claims and establish superhedging theorems for markets without and with short-sale prohibitions. 
The minimum superhedging costs are given by the suprema of discounted expected payoffs over the complete martingale and supermartingale dual families, respectively.

A rolling numerical experiment with synthetic put and call options on trees calibrated to S\&P 500 returns illustrates these results. Expanding the probability bounds widens the data-supported valuation intervals, while the reported mean upper values remain below the corresponding mean superhedging prices across the tested parameter values. 
The effect of short-sale prohibitions depends on the payoff: the mean put superhedging cost rises from $3.583\%$ to $13.648\%$ of spot, whereas the mean call cost remains $3.913\%$. Primal and dual values agree to numerical precision. 
In the out-of-sample evaluation, hedging deficits are observed only in windows with returns outside the calibrated range. For the put, the constrained cash-only strategy has a lower deficit frequency, accompanied by substantially greater initial funding and average surplus.
These findings distinguish the effect of probability uncertainty on valuation from the effect of trading constraints on hedging costs and performance.

The discrete framework provides a basis for pricing and hedging at specified trading dates, with probability uncertainty and portfolio constraints represented explicitly. Its finite-state structure permits backward linear programming and direct comparison of primal hedging costs with dual valuations.
These features make the discrete setting useful both for theoretical analysis and for numerical implementation. Future research may consider finer state grids, transaction costs, and continuous-time extensions, including the conditions under which discrete superhedges converge to admissible limiting strategies.


\end{document}